\documentclass[apj]{emulateapj}

\usepackage{comment}

\newcommand{\myemail}{meiert.grootes@mpi-hd.mpg.de}

\slugcomment{ Accepted by ApJ}

\shorttitle{Dust Opacity - Stellar Mass Surface Density}
\shortauthors{Grootes et al.}

\begin{document}


\title{GAMA/H-ATLAS: The Dust Opacity - Stellar Mass Surface Density Relation for Spiral Galaxies} 


\author{M.~W.~Grootes\altaffilmark{1,$\dagger$}, 
 R.~J.~Tuffs\altaffilmark{1}, 
 C.~C.~Popescu\altaffilmark{2}, 
 B.~Pastrav\altaffilmark{2}, 
  E.~Andrae\altaffilmark{1}, 
  M.~Gunawardhana\altaffilmark{3}, 
   L.~S.~Kelvin\altaffilmark{4,5}, 
   J.~Liske\altaffilmark{6}, 
   M.~Seibert\altaffilmark{7}, 
    E.~N.~Taylor\altaffilmark{3}, 
          A.~W.~Graham\altaffilmark{8}, 
          M.~Baes\altaffilmark{9},
           I.~K.~Baldry\altaffilmark{10}, 
           N.~Bourne\altaffilmark{11}, 
           S.~Brough\altaffilmark{12}, 
             A.~Cooray\altaffilmark{13}, 
             A.~Dariush\altaffilmark{14}, 
             G.~De~Zotti\altaffilmark{15,16}, 
              S.~P.~Driver\altaffilmark{4,5}, 
               L.~Dunne\altaffilmark{17}, 
               H.~Gomez\altaffilmark{18},
                A.~M.~Hopkins\altaffilmark{12}, 
                R.~Hopwood\altaffilmark{14,19}, 
                 M.~Jarvis \altaffilmark{20,21}, 
                 J.~Loveday \altaffilmark{22}, 
                 S.~Maddox\altaffilmark{17}, 
                 B.~F.~Madore\altaffilmark{7}, 
                 M.~J.~Micha{\l}owski \altaffilmark{23},
                 P.~Norberg\altaffilmark{24}, 
                 H.~R.~Parkinson\altaffilmark{23}, 
                  M.~Prescott\altaffilmark{10}, 
                  A.~S.~G.~Robotham\altaffilmark{4,5},  
                  D.~J.~B.~Smith\altaffilmark{20}, 
                  D.~Thomas\altaffilmark{25}, 
                  E.~Valiante\altaffilmark{18}} 

\altaffiltext{1}{Max-Planck-Institut f\"ur Kernphysik, Saupfercheckweg 1, 69117 Heidelberg, Germany}
\altaffiltext{$\dagger$}{\myemail}
\altaffiltext{2}{Jeremiah Horrocks Institute, University of Central Lancashire, Preston PR1 2HE, UK}
\altaffiltext{3}{Sydney Institute for Astronomy, School of Physics, University of Sydney, NSW 206, Australia}
\altaffiltext{4}{Scottish Universities' Physics Alliance (SUPA), School of Physics and Astronomy, University of St Andrews, North Haugh, St Andrews, KY16 9SS, UK}
\altaffiltext{5}{International Centre for Radio Astronomy Research (ICRAR), University of Western Australia, Stirling Highway Crawley, WA 6009, Australia}
\altaffiltext{6}{European Southern Observatory, Karl-Schwarzschild Str. 2, 85748, Garching, Germany}
\altaffiltext{7}{Observatories of the Carnegie Institution for Science, 813 Santa Barbara Street, Pasadena, CA 91101, USA}
\altaffiltext{8}{Centre for Astrophysics and Supercomputing, Swinburne University of Technology, Hawthorn, Victoria 3122, Australia} 
\altaffiltext{9}{Sterrenkundig Observatorium, Universiteit Gent, Krijgslaan 281 S9, B-9000 Gent, Belgium}
\altaffiltext{10}{Astrophysics Research Institute, Liverpool John Moores University, Twelve Quays House, Egerton Wharf, Birkenhead, CH41 1LD, UK}
\altaffiltext{11}{Centre for Astronomy and Particle Theory, The School of Physics \& Astronomy, Nottingham University, University Park Campus, Nottingham NG7 2RD, UK}
\altaffiltext{12}{Australian Astronomical Observatory, PO Box 296, Epping, NSW 1710, Australia}
\altaffiltext{13}{Department of Physics and Astronomy, University of California, Irvine, CA 92697, USA}
\altaffiltext{14}{Physics Department, Imperial College, Prince Consort Road, London, SW7 2AZ, UK}
\altaffiltext{15}{INAF-Osservatorio Astronomico di Padova, Vicolo Osservatorio 5, I-35122 Padova, Italy}
\altaffiltext{16}{SISSA, Via Bonomea 265, I-34136 Trieste, Italy}
\altaffiltext{17}{Department of Physics and Astronomy, University of Canterbury, Private Bag 4800, Christchurch, 8140, New Zealand} 
\altaffiltext{18}{School of Physics \& Astronomy, Cardiff University, Queen Buildings, The Parade, Cardiff, CF24 3AA, UK}
\altaffiltext{19}{Department of Physical Sciences, The Open University, Milton Keynes MK7 6AA, UK}
\altaffiltext{20}{Centre for Astrophysics Research, Science \& Technology Research Institute, University of Hertfordshire, Hatfield, Herts, AL10 9AB, UK}
\altaffiltext{21}{Physics Department University of the Western Cape, Cape Town, 7535, South Africa}
\altaffiltext{22}{Astronomy Centre, University of Sussex, Falmer, Brighton BN1 9QH; UK}
\altaffiltext{23}{Institute for Astronomy, University of Edinburgh, Royal Observatory, Blackford Hill, Edinburgh EH9 3HJ, UK}
\altaffiltext{24}{Institute for Computational Cosmology, Department of Physics, Durham University, Durham DH1 3LE, UK}
\altaffiltext{25}{Institute of Cosmology and Gravitation, Portsmouth University, Dennis Sciama Building, Portsmouth PO1 3FX, UK}


\begin{abstract}
We report the discovery of a well-defined correlation between B-band face-on central optical depth due to dust, $\tau^f_B$,
and the stellar mass surface density, $\mu_{*}$, of nearby ($ z \le 0.13$) spiral galaxies: $\mathrm{log}(\tau^{f}_{B}) = 1.12(\pm 0.11) \cdot \mathrm{log}\left(\frac{\mu_{*}}{\mathrm{M}_{\odot} \mathrm{kpc}^{-2}}\right)  - 8.6(\pm 0.8)$. This relation was derived from a sample of spiral galaxies taken from the Galaxy and Mass Assembly (GAMA) survey, which were 
detected in the FIR/submm in the \textit{Herschel}-ATLAS science demonstration phase field. Using a quantitative analysis of the NUV
attenuation-inclination relation for complete samples of GAMA spirals categorized according to stellar mass surface
density we demonstrate that this correlation can be used to statistically correct for dust attenuation
purely on the basis of optical photometry and S\'ersic-profile morphological fits. Considered together with previously established empirical relationships
of stellar mass to metallicity and gas mass, the near linearity and high
constant of proportionality of the $\tau^f_B\,-\,\mu_{*}$ relation
disfavors a stellar origin for the bulk of refractory grains in
spiral galaxies, instead being consistent with the existence of a
ubiquitous and very rapid mechanism for the growth of dust in the ISM.
 We use the $\tau^f_B\,-\,\mu_{*}$
relation in conjunction with the radiation transfer model for spiral
galaxies of Popescu \& Tuffs (2011) to derive intrinsic scaling
relations between specific star formation rate, stellar mass, and
stellar surface density, in which attenuation of the UV light
used for the measurement of star-formation rate is corrected 
on an object-to-object basis. A marked reduction in scatter
in these relations is achieved which we demonstrate is due
to correction of both the inclination-dependent and face-on
components of attenuation. Our results are consistent with a general
picture of spiral galaxies in which most of the submm emission
originates from grains residing
in translucent structures, exposed to 
UV in the diffuse interstellar radiation field.
\end{abstract}


\keywords{dust, extinction --- galaxies: fundamental parameters --- galaxies: ISM --- galaxies: spiral}



\section{Introduction}
\label{Introduction}
Broadband photometric imaging surveys in the UV/optical (e.g., SDSS \citep{ABAZAJIAN2009}, KiDS \citep{DEJONG2012}, EUCLID \citep{LAUREIJS2011}, GALEX MIS) are, and will continue to be, a main source of information 
from which the physical properties of galaxies must be deduced. 
It is, however, a well known issue that the UV/optical emission of galaxies 
is strongly attenuated by dust and that this attenuation should be taken into account \citep[e.g.][]{DRIVER2007,MASTERS2010}. 
This is particularly the case for late-type galaxies, which are usually much more gas-
and dust-rich than early-type galaxies (as recently re-confirmed using \textit{Herschel}-data by e.g., \citealt[][]{ROWLANDS2011, DARIUSH2011,BOURNE2011}). 
Furthermore, detailed imaging studies of dust emission in the Milky Way and nearby spiral galaxies
 \citep[e.g.][]{MOLINARI2010,BENDO2011,BRAINE2010,FRITZ2012} show that most of the dust is associated with large-scale structures in the neutral and molecular gas layers, which in turn causes the attenuation to depend heavily on disk inclination
\citep{TUFFS2004,PIERINI2004,DRIVER2007}.\newline

While it is generally agreed that the UV/optical emission of late-type galaxies must be corrected for dust attenuation,
it has proven to be a challenge to measure the opacities of the disks, and
various approaches exist. The most powerful method is to utilize infrared measurements of dust emission 
in combination with UV/optical data, since the attenuating
dust is heated by the UV/optical-emission it absorbs, and the bulk of this
energy is re-radiated longwards of 60 $\mu$m in the far-infrared (FIR) and
submm spectral range.
Approaches utilizing this UV/optical-FIR/submm information range from semi-empirical ones, such as the IRX absorption 
estimator \citep{MEURER1999,SEIBERT2005,JOHNSON2007}, via SED fitting using energy balance considerations \citep{DACUNHA2008,NOLL2009,SERRA2011}
, to 
radiation transfer modelling approaches, which explicitly calculate the UV/optical illumination of dust and the resulting FIR/submm-emission \citep[PT11 hereafter]{SILVA1998,POPESCU2000,  BIANCHI2000, GORDON2001, MISSELT2001, BIANCHI2008, BAES2010, BAES2011, MACLACHLAN2011, POPESCU2011}.
Unfortunately, such a coverage of the full UV/optical-FIR/submm SED is seldom,
or only incompletely available
for the population of spiral galaxies (despite their significant dust opacities) due to the scarcity of wide and sufficiently deep FIR surveys. Consequently, these methods can often only be
applied to more massive spiral galaxies and starbursts.\newline

In the absence of FIR data semi-empirical methods, based solely on UV/optical-data, such as the UV-spectral-slope
$\beta$ or the Balmer decrement are often applied \citep{CARDELLI1989,MEURER1999,CALZETTI2001, KONG2004,SEIBERT2005,WIJESINGHE2011}. These, however, often depend on
either multiple UV bands or spectroscopy, information which is also
often unavailable.\newline

In this paper we present a correlation between the B-band  face-on central dust-opacity $\tau_{B}^f$ and the stellar mass surface density $\mu_{*}$ 
of late-type, non-AGN galaxies, and demonstrate that this can be used in combination with the radiation transfer model of PT11 to statistically correct samples of
late-type galaxies for the attenuation due to dust using only broadband optical photometry.
The basis for this analysis is the overlap between the UV-optical-NIR/spectroscopic Galaxy and Mass Assembly survey (GAMA, \citealt{DRIVER2011}),
and the FIR/submm \textit{Herschel}-ATLAS (H-ATLAS, \citealt{EALES2010}) survey.
After briefly recapitulating some of the key concepts of the PT11 model (\S \ref{RTmodel}) and describing the data and samples employed (\S \ref{data}), we extract direct values of $\tau_{B}^f$
from the UV/optical-FIR-data 
for a sub-sample of local Universe late-type galaxies jointly measured by GAMA and H-ATLAS in the  H-ATLAS science demonstration phase (SDP) field  and use these to derive and calibrate the correlation between $\tau_{B}^f$ and $\mu_{*}$, as described in \S \ref{correlation}. 
In \S \ref{test} we show that the correlation correctly predicts the NUV attenuation-inclination relations for  complete sub-samples of GAMA late-type
galaxies characterized in terms of $\mu_{*}$, thus demonstrating its predictive power to correct samples of late-type galaxies without available FIR data for dust attenuation using only broadband optical photometry.
We then discuss our findings
in the context of the production/injection/survival of dust in spiral galaxies and the scaling relations between specific star formation rate, stellar mass, and stellar mass surface density in \S \ref{Discussion}, and summarize our results in \S \ref{summary}. 
We use AB magnitudes throughout this analysis, and adopt an $H_0 = 70$ km s$^{-1}$ Mpc$^{-1}$,
$\Omega_{M} = 0.3$, $\Omega_{\Lambda} = 0.7$ cosmology \citep{SPERGEL2003}.

\section{The Radiation Transfer model}
\label{RTmodel}
In this paper we quantitatively link the characteristics of the attenuation of starlight 
in spiral galaxies to the mass of dust in the galaxies as directly determined from the
FIR/submm integrated photometry. This approach mandates assumptions about
the spatial distribution of dust in the galaxies. Here we utilize the radiation transfer
model of \citet[][PT11 hereafter]{POPESCU2011}, which is applicable to a wide range
non-starburst, late-type galaxies. We refer the reader to PT11 as well as 
\citet{POPESCU2000,TUFFS2004} for a detailed description
of the model, its parameters, and the work done to test its performance.
Here we only supply a brief summary and detail its application to our data.\newline
The model consists of a bulge with an old stellar population and two exponential
disks describing the distribution of old and young stellar populations as well as
of diffuse dust located in these disks. This diffuse dust component can be
seen as representing dusty structures (such as Cirrus) with a substantial
projected surface filling factor. Emission from a dustless bulge is parameterized through
the inclusion of a bulge-to-disk ratio (B/D) to accommodate a range of
geometries along the Hubble sequence. In addition the model includes a
clumpy dust component with an embedded young stellar population
representing star-forming regions. The fraction of UV emission
escaping from the regions into the diffuse ISM is given by a factor $1 -F$
(fixed to $F=0.41$ for this analysis, following PT11).
This model, specifically the relative scale lengths and scale heights of the stars
and diffuse dust in the exponential disks, has been calibrated on and fixed to
the reproducible trends found in the local edge-on spiral galaxies 
analyzed in the radiation transfer analysis of \citet{XILOURIS1999}.
As such, the wavelength dependence of the scale lengths is also fixed.\newline

In the PT11 model the opacity of the disk at a given frequency and position can be expressed in terms of the central face-on optical depth of the combination of the two dust disks at a reference wavelength (PT11 use the B-band at 4430 {\AA}), $\tau^f_B$.
The value of $\tau^f_B$ can be expressed as:
 \begin{equation}
\tau^{f}_{B} = K\frac{M_{\mathrm{dust}}}{r_{s,d,B}^2}\;.
\label{taubf1}
\end{equation}
where $M_{\mathrm{dust}}$ is the total mass of dust in the galaxy, $r_{s,d,B}$ is the scale-length of the exponential disk in the B-band, and $K$ 
is a constant combining the fixed large-scale 
geometry and the spectral emissivity of the \citet{WEINGARTNER2001} model.
For the purposes of the work presented here the value of $\tau^f_B$ must be derived from observable
properties, hence Eq.~\ref{taubf1} must be re-expressed in terms of observational quantities.\newline

With the geometry of the model fixed, we express the physical scale length of the disk at the reference wavelength, $r_{s,d,B}$, using the corresponding angular size at the redshift at which the source is observed. We have chosen to determine this angular size in the $r$ band, which is less affected by the effects of dust attenuation than shorter wavelengths, while being less affected by noise than longer passbands, in particular the NIR, which may also suffer from systematic uncertainties (cf. e.g., \citet{TAYLOR2011}).\newline

The mass of dust $M_{\mathrm{dust}}$ is determined from the FIR/submm observations available from H-ATLAS. This data
extends longwards of $100\,\mu$m, thus predominantly sampling the emission by
cold dust in the galaxy and warranting the assumption that this range of the SED can be
reasonably approximated by a modified Planckian
$S_{\nu}(\nu)\,\sim\,\nu^{\beta}B(\nu,T)$ with $\beta=2$
(i.e the dust emission is not heavily affected by a warm dust component and/or stochastic heating processes broadening the peak of the SED). This
allows Eq.~\ref{taubf1} to be re-expressed as:

\begin{equation}
\tau^{f}_{B}  = A\frac{(1 + z)^{3 -\beta}}{B((1 + z)\nu_{\mathrm{250}},T_{0})}\frac{S_{\nu}(\nu_{250})}{\theta_{s,d,r}^2} \,,
\label{taubf2}
\end{equation} 

with $A=6.939\cdot 10^{-13}\,$~$\mathrm{arcsec}^2\,$~$\mathrm{J}\,$~$\mathrm{Jy}^{-1}\,$~$\mathrm{s}^{-1}\,$~$\mathrm{Hz}^{-1}\,$~$\mathrm{m}^{-2}\,$~$\mathrm{ster}^{-1}$, $\theta_{s,d,r}$ representing the $r$ band angular size corresponding to the disk scale length in arcsec, $S_{\nu}(\nu_{250})$ representing the observed mono-chromatic flux density at 250 $\mu$m in Jy, and $B(\nu,T)$ representing a Planckian with units of
$\mathrm{W}\,\mathrm{Hz}^{-1}\,\mathrm{m}^{-2}\,\mathrm{ster}^{-1}$, with a restframe temperature $T_{0}$. $T_{0}$,
$S_{\nu}(\nu_{250})$, and $\theta_{s,d,r}$ will be derived from measurements of
spatially integrated FIR/submm SEDs and optical morphologies in 
\S~\ref{correlation}.
The numerical value of $A$ has been calibrated using the detailed radiation transfer analysis results of the \citet{XILOURIS1999} galaxy sample. A detailed derivation of Eqs.~\ref{taubf1} and \ref{taubf2} and their link to the PT11 model together with a detailed description of the numerical calibration of $A$ is provided in appendix~\ref{AppendixTAUBF2}.\newline

The choice of using 250 $\mu$m is motivated by the tradeoff between using a measurement as far in the FIR/submm as possible, thus dominated by thermal emission of cold dust, and the sensitivity of the available data as discussed in \S \ref{data}.
The values of $\tau^f_B$ depend somewhat on the fitted restframe temperatures of the modified Planckian fits via the temperature dependence of the derived dust masses as shown in \S~\ref{Deriveopacities}; the typical uncertainty in the temperature of $\sim1\,$K corresponds to an uncertainty of the dust mass of $\sim15$\%.

\section{Data \& Samples}
\label{data}
The GAMA survey, currently comprising 11 band UV-NIR photometry (FUV, NUV, \textit{u}, \textit{g}, \textit{r}, \textit{i}, \textit{z}, Y, J, H, K), and $300 - 800$nm optical spectroscopy covering 
$\approx$ 144 deg$^2$ to $r_{petro} < 19.4$ mag with a spectroscopic completeness of $>$96 percent, provides a large, statistically complete, sample of local Universe galaxies with homogeneous photometry, spectroscopy, and ancillary data. This paper makes particular use of the stellar masses provided by \citet{TAYLOR2011} and single S\'ersic-profile fits performed by \citet{KELVIN2011}, providing morphological information. For a full description of the GAMA survey see \citet{DRIVER2011}, as well as \citet{BALDRY2010} for details of the input catalogue, \citet{ROBOTHAM2010} for details of the high-completeness tiling scheme, and \citet{HILL2011} together with Andrae et al. (2013, in prep.) 
for details of the photometry. 
 By design GAMA's coverage in the UV-optical-NIR is complemented in the FIR by the H-ATLAS survey \citep{EALES2010} using the PACS \citep{POGLITSCH2010} and SPIRE \citep{GRIFFIN2010} instruments on board
the \textit{Herschel Space Observatory} \citep{PILBRATT2010}. H-ATLAS achieves 5$\sigma$ point source sensitivities of 132, 126, 32, 26, and 45 mJy in the $100\,\mu \textrm{m}$, $160\,\mu \textrm{m}$, $250\,\mu \textrm{m}$, $350\,\mu \textrm{m}$, and $500\,\mu \textrm{m}$ channels respectively. The details of the SPIRE and PACS map-making process are described in \citet{PASCALE2011} and \citet{IBAR2010}, while the catalogues are described in \citet{RIGBY2010}. The FIR data used in this analysis are taken from the catalogue of 
H-ATLAS sources matched to optical SDSS/GAMA sources using a likelihood-ratio method \citep{SMITH2011}, with a required reliability  of $>$80$\%$. This catalogue is defined by the requirement of 5-$\sigma$ detections at 250$\,\mu$m.\newline

\begin{figure*}
\plotone{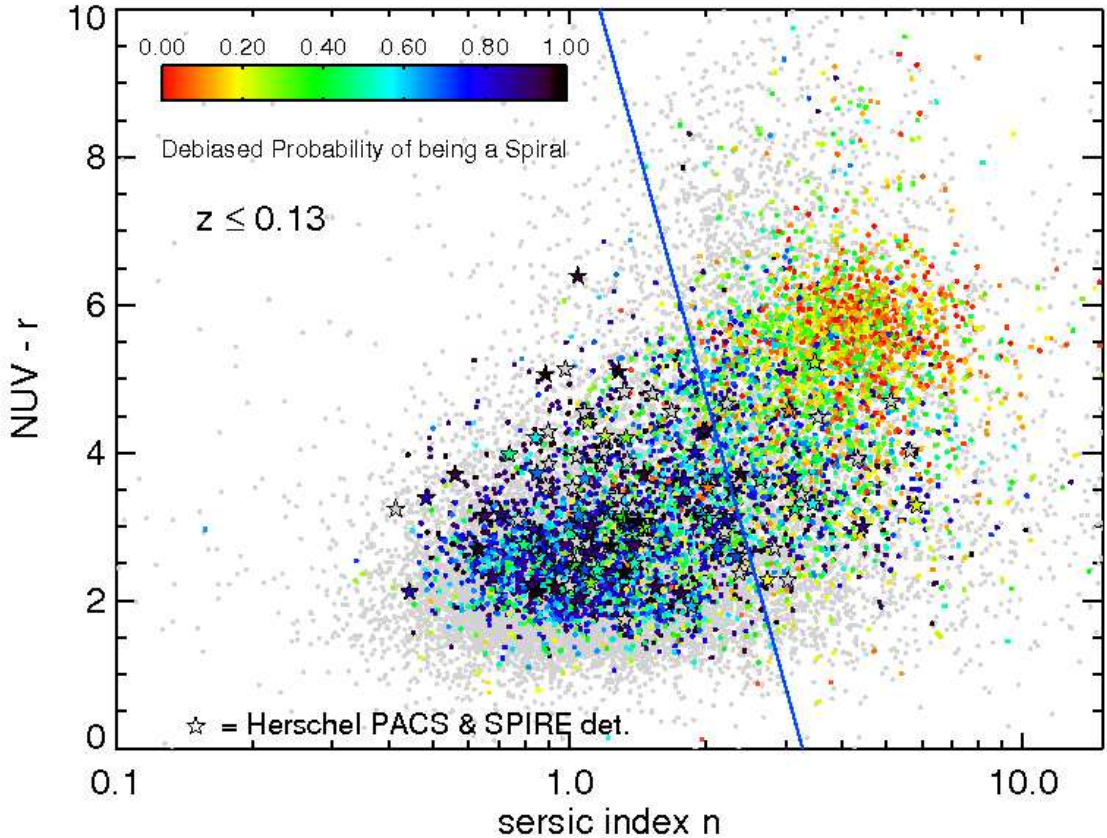}
\caption{NUV-\textit{r} vs. $n_r$ for non-AGN GAMA galaxies ($z \le 0.13$) with NUV and \textit{r} band detections. The color-coded sub-sample represents the overlap with GALAXY ZOO DR1 sources with debiased morphological classifications \citep{LINTOTT2008,BAMFORD2009} 
with the color representing the probability that the source is a spiral. The blue line represents an automated morphological selection designed to reliably select late-type galaxies with minimal contamination, calibrated on the GALAXY ZOO data (see GR12 for details)
. Overplotted as stars are the sources detected in the H-ATLAS SDP field with SPIRE and PACS, with those in \textit{OPTICAL+FIR} located left of the divide.}
\label{figmorphsel}
\end{figure*}

In the context of this work it is critical to select normal,
i.e. non-starburst, non-AGN, late-type galaxies as reliably as possible, as the radiation
transfer model employed is designed to treat, and is calibrated on, such sources, and is therefore only
accurate for these. We
first select those GAMA sources which have NUV and \textit{r} band detections,
reliable spectroscopic redshifts, and which have not been identified as AGN
following the prescription of \citet{KEWLEY2006}\footnote{Almost all these
  sources have available S\'ersic parameters and stellar mass estimates. Those
  that do not are excluded from the sample}, limiting ourselves to a local
Universe sample with $z \le 0.13$ to avoid evolutionary effects\footnote{This
  limitation in redshift also is conducive to our confidence in the derived
  values of the S\'ersic parameter n, as
meaningful morphological fits become more difficult at larger redshifts.}.
From these 14998 galaxies we select a population of probable spiral galaxies in the NUV-\textit{r} vs.  \textit{r}-band S\'ersic index $n_r$ plane \citep{DRIVER2012,KELVIN2011} as those with
\begin{equation}
(\mathrm{NUV} - r) \le -22.35\, \mathrm{log}(n_r) + 11.5 \,,
\label{morphsel}
\end{equation}
resulting in a subset of 11236 galaxies which we will refer to as the \textit{OPTICAL} sample.\newline 
The separator defined by Eq. \ref{morphsel} corresponds to the blue line in Fig.\ref{figmorphsel}. It was determined from GAMA sources with debiased visual morphological classifications from GALAXY ZOO DR1 \citep{LINTOTT2008,BAMFORD2009}, and is designed to maximize the number of GALAXY ZOO spirals correctly classified while simultaneously minimizing the contamination of the spiral sample due to GALAXY ZOO ellipticals based on a figure of merit maximization approach 
. As such, it is a conservative classification and will exclude some real spirals. We find a contamination of the resulting spiral sample by reliably classified ellipticals of 5-10\% .\newline
\newline
From \textit{OPTICAL} we select the subset of sources with $> 3 \sigma$ detections in the 100$\mu$m and/or 160$\mu$m channel in addition to at least a 5$\,\sigma$ detection in the 250$\mu$m channel (i.e. from the H-ATLAS SDP field where both SPIRE and PACS catalogues are currently available).
We then calculate values of $\tau^{f}_{B}$ for these 79 sources as described in \S \ref{correlation} excluding two with $\tau^{f}_{B} > 30$ (which 
we take to be indicative of starburst systems).
Finally we exclude three sources which have been visually classified as early-type galaxies by \citet{ROWLANDS2011}, resulting in a sample of 74 late-type galaxies with detections in at least two FIR bands, referred to as the \textit{OPTICAL+FIR} sample.\newline

\section{The opacity - surface density correlation}
\label{correlation}
\subsection{Deriving opacities}\label{Deriveopacities}
For each of the galaxies in the \textit{OPTICAL+FIR} sample the disk opacity was calculated from Eq.~\ref{taubf2} using knowledge of $T_{0}$ and $S_{250}$ (derived from \textit{Herschel} data), and the \textit{r}-band angular exponential disk scale $\theta_{s,d,r}$ (derived from $\theta_{e,ss,r}$ the \textit{r}-band single S\'ersic effective size in arcsec (i.e, the half-light radius) catalogued by \citealt{KELVIN2011}).\newline

To derive $T_{0}$ from the \textit{Herschel} data we fit
isothermal modified Planckians ($\beta=2$) to all available data points. The requirement of a detection at $160\,\mu$m or shortwards allows the spectral peak of the dust emission to be well constrained. We find a median value of $23.2\,$K for $T_0$. The value of $T_0$ is almost independent of the wavelengths at which the \textit{Herschel} data are obtained, as the median temperature of sources with only a $160\,\mu$m PACS detection is $22.58\,$K, while that of sources with a PACS detection only at $100\,\mu$m is $23.58\,$K. The median temperature of sources with PACS detections at both $100\,\mu$m and $160\,\mu$m is $23.35\,$K. The difference in median dust temperature of galaxies between the $100\,\mu$m-only and the $160\,\mu$m-only sample corresponds to an uncertainty in the dust mass of $\approx 15$\%.
The requirement of a datapoint at $160\,\mu$m or shortward does not appear to induce a strong bias towards warmer sources, since the median temperature of our sample is consistent with the mean value of $22.7\pm2.9\,$K for blue galaxies with $3\times10^9<M_{*}<3\times10^{11} M_{\odot}$ (roughly comparable to our sample) found by \citet{BOURNE2011} using a stacking analysis of H-ATLAS data on optically selected galaxies.\newline

Overall, we believe that the isothermal model constrained by SPIRE data at $\lambda \ge 250\,\mu$m and a PACS data point at 100 and/or $160\,\mu$m represents a robust
method of determining dust masses using minimal assumptions, due to the decrease in the uncertainty of both temperature and amplitude arising from a data point constraining the peak of the dust emission, and because the wavelength coverage ($\lambda > 100\,\mu$m) misses any significant emission arising from warm dust in SF regions or from stochastically-heated small grains in the diffuse ISM \citep[e.g.,][]{POPESCU2000, GORDON2001,MISSELT2001,GALLIANO2003,GALLIANO2005}.\newline

Using Eq.~\ref{taubf2} to determine $\tau^f_B$ requires knowledge of the angular size corresponding to the disk-scale length in the $r$ band. 
The relation between the observable single S\'ersic effective size and the disk scale-length of a spiral galaxy, however, is non-trivially influenced by the relative fraction of emission from the bulge and the disk as well as by dust present in the galaxy, with the former causing the ratio between $\theta_{e,ss,r}$ and $\theta_{s,d,r}$ to decline, while the latter tends to cause sizes to be overestimated, increasing the ratio. Pastrav et al. (in prep.) have investigated the combined dependencies of the ratio between $\theta_{e,ss,r}$ and $\theta_{s,d,r}$ on bulge-to-disk ratio, dust opacity, inclination and wavelength, and provide their results in tabulated form. In this work we have self-consistently determined $\tau^f_B$ for the \textit{OPTICAL+FIR} sample using Eq.~\ref{taubf2} and the results of Pastrav et al. as detailed in appendix~\ref{AppendixSizes}. In doing so, we have used a bulge-to-disk ratio of $B/D=0.33$ found to be representative of the generally earlier type, more massive spirals \citep{GRAHAM2008}, such as those in the \textit{OPTICAL+FIR} sample. We caution, that this use of an average value of $B/D$ will introduce uncertainty into the derived values of $\tau^f_B$ as shown in appendix~\ref{AppendixSizes}, and will revisit our results when reliable bulge+disk decompositions based on higher resolution imaging of these objects becomes available.\newline

\subsection{Deriving stellar mass surface densities}\label{Derivemustar}
We compute the stellar mass surface density $\mu_{*}$ using the physical radius $r_{e,ss,r}$ corresponding to the single S\'ersic effective radius in arcsec provided by \citet{KELVIN2011} and the GAMA stellar masses $M_{*}$ from \citet{TAYLOR2011} as
\begin{equation}
\mu_{*} = \frac{M_{*}}{2 \pi r^2_{e,ss,r}} = \frac{M_{*}}{2\pi D^{2}_{A}(z)\theta^{2}_{e,ss,r}}\;\,,
\label{eqmustar}
\end{equation}
where $D^{2}_{A}(z)$ is the angular diameter distance corresponding to the redshift $z$. 
We note that the stellar masses predicted by \citeauthor{TAYLOR2011} 
incorporate a single fixed prediction of the reddening and
attenuation due to dust derived from \citet{CALZETTI2000}.
Thus, expected systematic variations in reddening and attenuation
with inclination, disk opacity and bulge-to-disk ratio
are not taken into account.
However, as discussed by \citeauthor{TAYLOR2011}
(see also Fig.~12 of \citealt{DRIVER2007}) the resulting shifts in
estimated stellar mass are much smaller than the individual effects
on color and luminosity, and should not significantly affect the
relation between disk opacity and stellar mass surface density derived in
this paper. Taking this, and other effects into account, the
typical uncertainty in the stellar mass estimated by \citeauthor{TAYLOR2011}
is $\sim\,0.1\,$dex.\newline 

We also note that \citet{TAYLOR2011} make use of a \citet{CHABRIER2003} IMF and the \citet{BRUZUAL2003} stellar population library, and that hence,
any systematic variations due to the choice of IMF or the stellar population library are not taken into account.

\subsection{The relation between opacity and stellar mass surface density}\label{tauvmucorr}
\begin{figure*}
\plotone{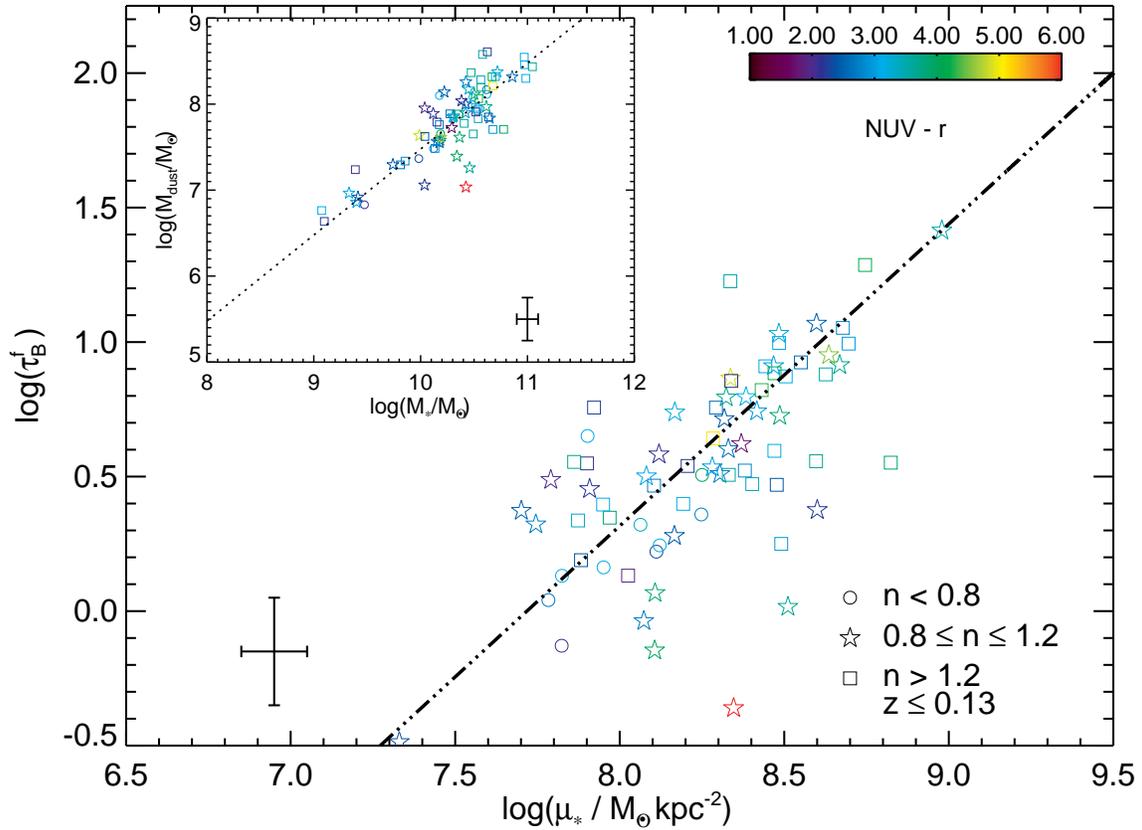}
\caption{B-band face-on central optical depth $\tau^{f}_{B}$ vs.  stellar mass surface density $\mu_{*}$ for \textit{OPTICAL+FIR} galaxies. Symbols are coded according to $n$ and NUV-\textit{r} color (see figure). 
The dash-dotted line represents the best-fit (Eq. \ref{correleq}). The median uncertainties in $\tau^{f}_{B}$ and $\mu_{*}$ are depicted at bottom left. 
The inset depicts the dust mass (derived from $\tau^f_b$ using Eqs.~\ref{taubf1} and \ref{taubf2}) as a function of stellar mass. The dotted line represents a reference value with a slope of unity and an offset corresponding to $M_{\mathrm{dust}}/M_{*} = 0.003$. Median errors are depicted bottom right.
}
\label{Fig_corr}
\end{figure*}

$\tau^f_{B}$ is plotted against $\mu_{*}$ for the \textit{OPTICAL+FIR} sample in Fig.~\ref{Fig_corr}. The data points are shown as symbols according to their S\'ersic index, with the color corresponding to the NUV-\textit{r} color.
Using a linear regression analysis taking the uncertainties in both $\mu_{*}$ and $\tau^{f}_{B}$ into account we find a power-law correlation between the two with $\chi ^{2} /N_{\mathrm{DOF}} = 0.97$ ($N_{\mathrm{DOF}} = 72$) as
\begin{equation}
\mathrm{log}(\tau^{f}_{B}) = 1.12(\pm 0.11) \cdot \mathrm{log}\left(\frac{\mu_{*}}{\mathrm{M}_{\odot} \mathrm{kpc}^{-2}}\right)  - 8.6(\pm 0.8)
\label{correleq}
\end{equation}
depicted by the dash-dotted line in Fig. \ref{Fig_corr}. The errors represent the 1-$\sigma$ uncertainties in the regression analysis. \newline
The correlation is tightest for sources with NUV-\textit{r} colors of $\sim 3$ with a slight 
increase in scatter for bluer and redder colors. This increase is likely due in part to the assumed B/D ratio in the determination of $\tau^f_B$ as detailed in appendix~\ref{AppendixSizes}, but may also represent a larger range in opacities for bluer possibly more irregular system and redder systems which may appear red either due to dust, or because they are more passive systems.
There is also evidence for such a population of passive spirals, i.e, spirals with low $\tau^f_B$ high $\mu_{*}$, as presented by \citet{ROWLANDS2011}.\newline
We find that 50\% of the sample lie within $0.14\,$dex of the correlation ( $\Delta_{s,0.5} \mathrm{log}(\tau^{f}_{B}) = 0.14$), comparable to the median measurement error for $\mathrm{log}(\tau^{f}_{B})$.
Thus it is possible that a large fraction of the visible scatter is due to measurement uncertainties.\newline

\subsubsection{Range of applicability and limitations of the relation}
\begin{figure*}
\plotone{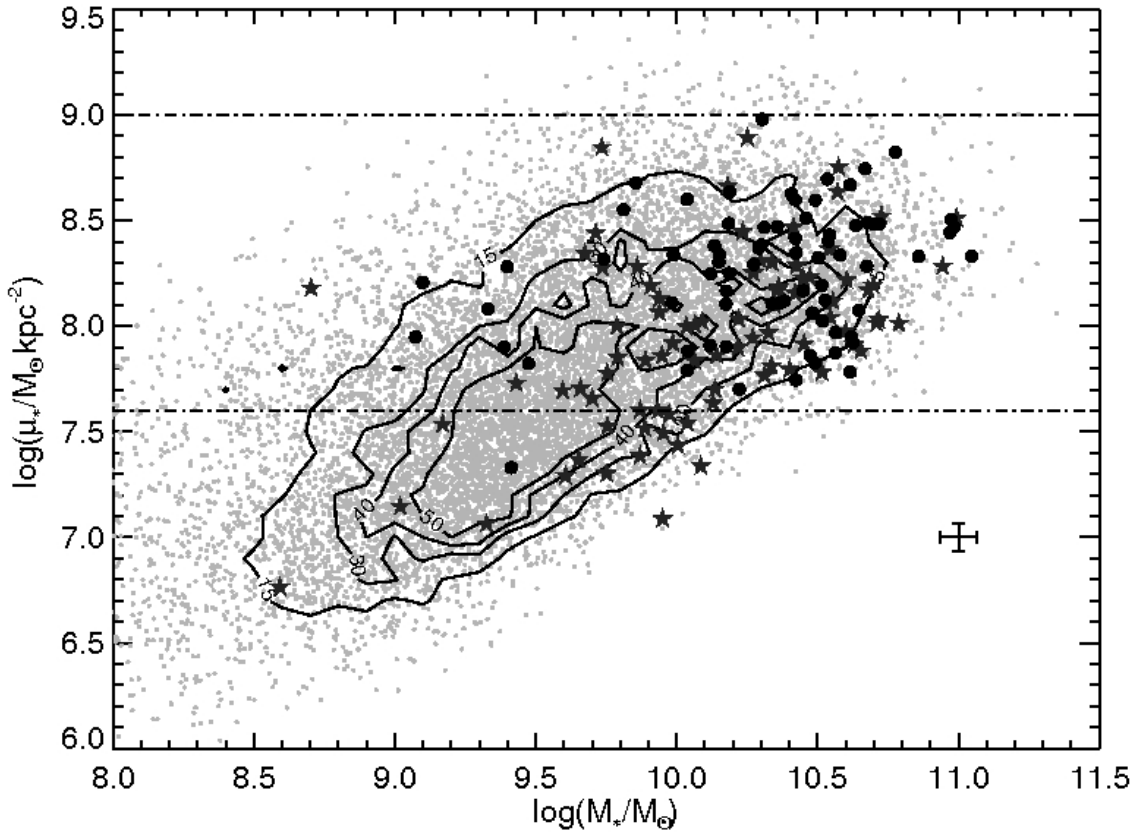}
\caption{$\mu_{*}$ as a function of $M_{*}$ for \textit{OPTICAL} (gray) galaxies with isodensity contours. Overplotted are the \textit{OPTICAL+FIR} sources (circles) and those sources which fulfill the morphological selection but only have SPIRE detections (stars). Dash-dotted lines indicate the range in $\mu_{*}$ for which the correlation has been calibrated. The median errors on both properties are shown at bottom right.}
\label{Fig_SMSMSD}
\end{figure*}

In order to understand the range of applicability of the correlation shown in
Fig.~\ref{Fig_corr} and given by Eq.~\ref{correleq}, as well as to identify
possible biases caused by the use of a FIR-selected sample in deriving this
result, we have overplotted the distribution of the \textit{OPTICAL+FIR} sample
in the $\mu_{*}$ vs. $M_{*}$ plane on that of the \textit{OPTICAL} sample in
Fig.~\ref{Fig_SMSMSD}. The \textit{OPTICAL+FIR} sample covers a range of $7.6<
\mathrm{log}(\mu_{*}) < 9.0$ in $\mu_{*}$ more or less uniformly and can be
deemed applicable in this range. Fig.~\ref{Fig_SMSMSD}, however, also clearly
shows that the \textit{OPTICAL+FIR} sample is strongly biased towards more
massive sources, as shown by the positions of the purple stars. This bias
arises from the the fact that the \textit{OPTICAL+FIR} is defined by the
sensitivity of the \textit{Herschel} instruments. In spite of this clear bias
in stellar mass, however, the sample does contain sources which provide a
tentative coverage of the entire stellar mass range corresponding to the range
in $\mu_{*}$ as seen in the \textit{OPTICAL} sample. As discussed in
\S~\ref{test}, this bias in stellar mass does not affect the applicability of
Eq.~\ref{correleq} to large samples of galaxies as constituted by the
\textit{OPTICAL} sample.\newline

The range of $7.6< \mathrm{log}(\mu_{*}) < 9.0$ for which Eq.~\ref{correleq}
may be deemed applicable only provides a complete sample of galaxy stellar mass
above $10^{9.5}\,M_{\odot}$. Thus, if the correlation is to be applied to
samples of galaxies which need to be complete in stellar mass, these should be
accordingly limited until the correlation can be calibrated for lower values of
$\mu_{*}$ (corresponding also to lower values of $M_{*}$).\newline

Fig.~\ref{Fig_SMSMSD} also shows that $\mu_{*}$ appears to be loosely correlated with $M_{*}$, with higher mass galaxies having larger values of $\mu_{*}$. This raises the question of whether the $\tau^f_{B}$ - $\mu_{*}$ correlation shown in Fig.~\ref{Fig_corr} is actually a relation between $\tau^f_{B}$ and $M_{*}$. Using a linear partial correlation analysis of $X=\mathrm{log}(M_{*})$, $Y=\mathrm{log}(\mu_{*})$, and $Z=\mathrm{log}(\tau^{f}_B)$ on the \textit{OPTICAL+FIR} sample we obtain the partial correlation coefficients $r_{XY,Z} = 0.271$, $r_{XZ,Y}=-0.047$, and $r_{YZ,X} = 0.611$, however, implying that $\mu_{*}$ is indeed the dominant factor in determining $\tau^{f}_B$.\newline

Furthermore, as Eq.~\ref{correleq} presents a relation between two properties
which are both inversely proportional to an area, we must ask ourselves whether
the result is actually a spurious correlation due to noise in the size
measurements. Due to the moderate redshift limit of $z \leq 0.13$, however, the
uncertainties on the size determination are much smaller than the range in
sizes found for a given value of $M_{*}$, showing that the spread in values of
$\mu_{*}$ is mainly intrinsic.\newline

Finally, we emphasize that the quantitative accuracy of the relation given by Eq.~\ref{correleq} depends on the applicability of the large-scale geometry of the exponential dust disks as calibrated in PT11 to the range of late-type galaxies with $7.6< \mathrm{log}(\mu_{*}) < 9.0$.\newline

\subsection{The dust mass-to-stellar mass relation}
\label{MDSM}
It is clear that $\tau^f_B$ is akin to a surface density and requires measurements of both a galaxy's stellar mass and size to facilitate it's estimation.    
Nevertheless, a major underlying physical driver for the result presented by Eq.~\ref{correleq} is a roughly linear correlation between the mass of  dust and stars in late-type galaxies of the \textit{OPTICAL+FIR} sample. This is shown in the inset of Fig.~\ref{Fig_corr}, where dust masses (derived from the values of $\tau^f_B$ using Eq.~\ref{taubf1}) are plotted against stellar masses from \citet{TAYLOR2011}. The dotted line depicts a slope of unity with a dust to stellar mass fraction of $3\cdot 10^{-3}$ as a reference value.\newline

\subsubsection{Comparison with other \textit{Herschel} results}
Several previous works have provided data on the dust-to-stellar mass ratio for different samples of galaxies, allowing quantitative comparisons with our results. \citet{SKIBBA2011} present stellar and dust masses for the galaxies in the Herschel KINGFISH survey \citep{KENNICUTT2011}. For a sample of spiral galaxies of type Sa and later with $M_{*} > 10^9 \mathrm{M}_{\odot}$, comparable to our sample, one finds an average dust-to-stellar mass ratio of $-3.02\pm0.5$ (derived from Table~1 of \citet{SKIBBA2011} ), comparable within errors to our result. 
Furthermore the dust-to-stellar mass ratio inferred by our data is comparable within errors to that found for spiral galaxies in the Herschel Reference Survey \citep[HRS,][]{BOSELLI2010}  by \citet{CORTESE2012}, as shown in Fig.~5 of \citet{CORTESE2012} for individual morphological types and in Fig.~9 of \citet{MSMITH2011} for all spiral galaxies in the HRS. The agreement is particularly good for the dust-to-stellar mass ratios derived for earlier-type spirals which, on average, are more massive and are likely to be more directly comparable to our sample \citep[][their Fig.~5]{CORTESE2012}. In addition Fig.~5 of \citet{CORTESE2012} also shows that the dust-to-stellar mass ratio is nearly constant for galaxies with morphological type Sa and later, especially for galaxies with HI-deficiencies generally indicative of residing in environments comparable to our sample (i.e., not being  members of massive clusters).  

Finally we find that the dust-to-stellar mass ratio of $\sim 3\cdot 10^{-3}$, is also in
general agreement with that derived by \citet{DUNNE2011} ($ \sim 2\times 10^{-3}$)
for low redshift galaxies using all H-ATLAS SDP field sources, although we note
that these authors have employed a different calibration of FIR/submm dust
emissivity. \newline
 
The roughly linear slope of the relation between dust and stellar mass in the \textit{OPTICAL+FIR} sample also generally agrees well with the data for late-type HRS galaxies plotted in Fig.~8 of \citet[][]{MSMITH2011}. The data show a slope which is slightly sub-linear over a large range in stellar mass extending down to below $10^9\,M_{\odot}$. At galaxy stellar masses above $10^{9.5}\,M_{\odot}$ (more similar to our \textit{OPTICAL+FIR} sample), however, the data exhibit a slope which is considerably closer to unity. The results of \citet{BOURNE2011}, who find a correlation between dust and stellar mass based on a stacking analysis of optically selected sources, display similar properties, with the relation between dust and stellar mass steepening with increasing stellar mass (their Fig.~16). We note that \citet{BOURNE2011} do not apply a morphological classification, but rather categorize their sample into blue, green, and red bins according $g-r$ color which is likely to place some of the dusty edge-on
spirals included in our sample in the green or even red bin in which their data displays a steeper, more linear slope.\newline

\section{Testing the predictive power of the stellar mass surface density - opacity relation}
\label{test}
The $\tau^f_B$ - $\mu_{*}$ relation as formulated in Eq.~\ref{correleq} is fundamentally an empirical result, linking the total dust mass per directly measured area to a direct measure of the stellar mass per unit area. Thus, the relation is largely independent of detailed assumptions about the distribution of dust within the galaxy disks. Nevertheless, since we have formulated the previous analysis in the terms of the $\tau^f_B$ parameter of the radiation transfer model of PT11, we are in a position to directly and independently test the physical consistency and predictive power of Eq.~\ref{correleq} using any observable effect which is predicted to be a function of  $\tau^f_B$ by the PT11 model. To this end we analyze here two quantities which are dependent on the amount and distribution of dust in galaxies, using a much bigger (and largely disjunct in terms of FIR detections) sample than used for the calibration of the $\tau^f_B$ - $\mu_{*}$ relation, i.e., the \textit{OPTICAL} sample. These quantities are the inclination dependence of attenuation of NUV emission from late-type galaxies (considered in \S~\ref{attincrel}) and the scatter about the well known scaling relation between the specific star formation rate $\psi_*$ and the stellar mass $M_*$ (considered in \S~\ref{scatterpsiSM}). We will show that even though the $\tau^f_B$ - $\mu_*$ relation is calibrated on a very limited portion of the overall population of galaxies it is applicable to the general population of spiral galaxies with $7.6< \mathrm{log}(\mu_{*}) < 9.0$. \newline

\subsection{Deriving attenuation corrections}\label{deriveatcor}
The radiation transfer model presented in \citet{TUFFS2004} and PT11 allows the inclination-dependent attenuation of a spiral galaxy to be calculated
for known values of $\tau^{f}_{B}$, disk inclination $i_d$, escape fraction $F$, and bulge-to-disk-ratio $B/D$ using Eqs.~17 \& 18 from \citet{TUFFS2004} and the model predictions of attenuation coefficients tabulated in PT11\footnote{The requisite data specifying attenuation as a function of different wavelengths are available in electronic form at the CDS via anonymous ftp to cdsarc.u-strasbg.fr (130.79.128.5) or via http://cdsarc.u-strasbg.fr/viz-bin/qcat?J/A+A/527/A109.}.
For the UV the value of $B/D$ is of negligible importance as the UV emission is almost entirely produced in the disk, even for early-type spirals, and we have assumed $F=0.41$ throughout as calibrated in PT11. Values of $i_d$ and $\tau^f_B$ leading to an object-by-object estimate of attenuation are found as in the following. 

\subsubsection{Deriving inclinations} \label{Deriveinc}
The \textit{OPTICAL} galaxy inclinations are calculated from the \textit{r}-band single S\'ersic fit axis-ratios of \citet{KELVIN2011} as
$i_d = \arccos((b/a) _{ss})$, where $(b/a)_{ss}$ represents the single S\'ersic axis-ratio in the $r$ band. These inclinations are then corrected for the effects of finite disk-thickness as detailed in \S~3 of 
\citet{DRIVER2007}, with an assumed intrinsic ratio of scale-height to semi-major axis of 0.12.\newline

\subsubsection{Calculating attenuation-corrected magnitudes}\label{cormags}
Using the  catalogued stellar masses \citep{TAYLOR2011} and the measured values of $\theta_{e,r}$ \citep{KELVIN2011} together with Eq.~\ref{correleq} we estimate the values of $\tau^f_{B}$ for the entire \textit{OPTICAL} sample. \newline
We then correct the NUV absolute magnitudes using the radiation transfer model (\citealt{TUFFS2004}, utilizing the aforementioned tables of attenuation coefficients in PT11) together with the disk inclination $i_d$ and $\tau^f_{B}$.\newline

\subsection{The attenuation-inclination relation in the NUV}\label{attincrel}
Previous work \citep[e.g,][]{TUFFS2004,DRIVER2007,MASTERS2010} 
has predicted and shown that the attenuation of UV/optical-emission in spiral galaxies is a strong function 
of inclination,
with this effect being particularly pronounced at shorter wavelengths, thus
severely influencing, for example, UV-based tracers of star-formation. 
This attenuation-inclination relation implies that the median observed absolute magnitude 
of members of a given late-type galaxy population should increase as a function
of inclination.\newline

Here, given measurements of the inclinations, we use the attenuation-inclination relation to test 
the predictive power and physical consistency of Eq. \ref{correleq} by
calculating the intrinsic absolute NUV magnitudes $\mathrm{M}_{\mathrm{NUV}}$, corrected for attenuation as detailed in \S~\ref{deriveatcor}.
On an object-by-object basis the values of $M_{NUV}$ will display scatter, at the very least due to an intrinsic spread in the galaxies' physical quantities. However, the median of an optically selected sample should no longer display an inclination-dependence after correction, if the transfer of UV radiation in galaxies is adequately described by the PT11 model, and the $\tau^f_B$ - $\mu_{*}$ correlation given by Eq.~\ref{correleq} is representative of the late-type galaxy population as a whole (in the according range of $\mu_{*}$).
In particular, given the bias towards massive galaxies in the \textit{OPTICAL+FIR}, the applicability of Eq.~\ref{correleq} to the galaxy population as a whole is by no means obvious.\newline

\begin{figure*}
\plotone{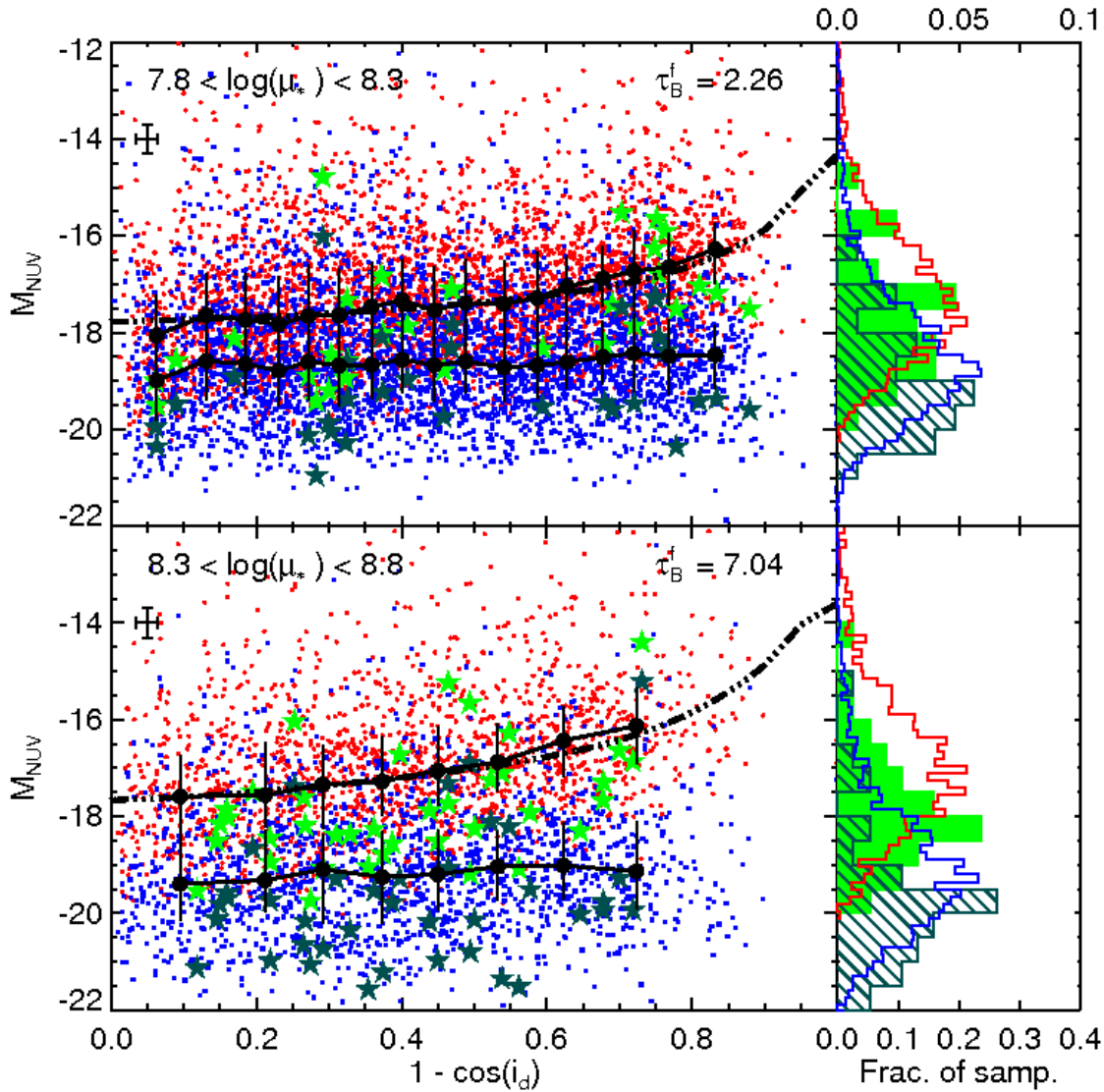}
\caption{Uncorrected (red circles) and corrected (blue squares) values of $\mathrm{M}_{\mathrm{NUV}}$ vs. $1-\mathrm{cos}(i_d)$ for two sub-samples defined in $\mu_{*}$ as stated in the figure. The samples include all values of $M_{*}$ present in the relevant ranges of $\mu_{*}$ of the \textit{OPTICAL} sample. Sources are binned in equal numbers (250) with the bin-wise median $\mathrm{M}_{NUV}$ and $1-\mathrm{cos}(i_d)$ depicted by solid black circles connected by solid lines, and error bars indicating the quartile boundaries. \textit{OPTICAL+FIR}-sources are overplotted in green(uncorrected) and gray(corrected). The black dash-dotted line traces the inclination dependence predicted by the PT11 radiation transfer model 
for a fiducial galaxy with sample-defined median $\tau^{f}_{B}$ (2.26 resp. 7.04, see figure), and median intrinsic $\mathrm{M}_{NUV}$, defined by the corrected sample.  Median random errors are shown shown top left, however, we expect increasing systematic errors in the determination of disk inclination at higher inclinations (see text). The histograms show the collapsed distributions in $\mathrm{M}_{\mathrm{NUV}}$ for the \textit{OPTICAL} sample before and after corrections for dust attenuation (red and blue histograms respectively, with upper ordinate) and for the \textit{OPTICAL+FIR} sample also before and after correction (green and blue hatched histograms respectively, with lower ordinate).  }
\label{FIGtest}
\end{figure*}

Fig.~\ref{FIGtest} shows the distributions of corrected and uncorrected absolute NUV magnitude $M_{NUV}$ as a function of inclination given as $1 - \mathrm{cos}(i_d)$, for two sub-samples of \textit{OPTICAL} defined by distinct ranges of $\mu_{*}$, thus corresponding to very different mean values of $\tau^f_{B}$.  The samples are drawn from the range of observed stellar mass surface density $\mu_{*}$ for which Eq.~\ref{correleq} is applicable ($7.6< \mathrm{log}(\mu_{*}) < 9.0$; see \S \ref{tauvmucorr} and Fig.~\ref{Fig_SMSMSD}), and cover the complete range of available galaxy stellar masses. The ranges in $\mu_{*}$ have been chosen to ensure that the samples are not affected by biases due to noise scattering sources into or out of the range in $\mu_{*}$ for which  Eq.~\ref{correleq} has been calibrated.
We find a median value of $\tau^f_{B,int} = 2.26$ for the galaxy sample with $7.8<\mathrm{log}(\mu_{*})<8.3$ and a value of $\tau^f_{B} = 7.04$ for the sample with $8.3<\mathrm{log}(\mu_{*})<8.8$. The median value for both samples combined is $\tau^f_{B} = 3.10$, while that for the entire \textit{OPTICAL} sample with $7.6 \le \mathrm{log}(\mu_{*}) \le 9.0$, the range for which Eq.~\ref{correleq} has been calibrated, is $\tau^f_{B} = 2.42$.\newline

As can be clearly seen the uncorrected samples (red points in Fig.~\ref{FIGtest}) display a clear inclination-dependent dimming of their magnitudes, with the median magnitude increasing (i.e. dimming) from the face-on case ($1 - \mathrm{cos}(i_d) = 0$) to the edge-on case ($1 - \mathrm{cos}(i_d) = 1$).
In both ranges of $\mu_{*}$ the attenuation-corrected values of $M_{NUV}$, derived as described in \S \ref{cormags}, are shown in blue.
The corrected values of $M_{NUV}$ no longer display a dependence on inclination, indicating that the correlation found using the \textit{OPTICAL+FIR} sample is consistent with the independent observable presented by the attenuation-inclination relation, and with the radiation transfer model of PT11. This is also consistent with the supposition that the bias towards massive/bright sources in the \textit{OPTICAL+FIR} sample, discussed in \S \ref{tauvmucorr} and shown in Fig.~\ref{Fig_SMSMSD}\footnote{The bias of the \textit{OPTICAL+FIR} sample towards bright sources is also visible in Fig.~\ref{FIGtest} where the uncorrected (green) and corrected (gray) values of $M_{NUV}$ for the galaxies in the \textit{OPTICAL+FIR} sample in the appropriate range in $\mu_{*}$ are overplotted and predominantly lie at the bright edge of the distribution.}
 does not affect the correlation's applicability to the much larger
 \textit{OPTICAL} sample.\newline

 These conclusions are reinforced on a quantitative level by the 
agreement between the observed median distribution of the uncorrected samples and the predicted inclination dependence of a fiducial galaxy with $\tau^{f}_{B}$ corresponding to the median of the
sample, and $\mathrm{M}_{\mathrm{NUV}}$ corresponding to the median of the corrected bin-wise median $\mathrm{M}_{\mathrm{NUV}}$, depicted by the dash-dotted line in Fig.~\ref{FIGtest}.
In addition to the predicted dependence of attenuation on inclination, the difference in gradient of the attenuation as a function of inclination predicted by PT11 for $\tau^f_{B} = 2.26$ (corresponding to the range of  $7.8<\mathrm{log}(\mu_{*})<8.3$) and $\tau^f_{B} = 7.04$ (corresponding to the range of  $8.3<\mathrm{log}(\mu_{*})<8.8$) is also shown in the data.\newline

A large uncertainty in quantitative interpretations of the attenuation-inclination relation such as these arises from the difficulty of correctly
classifying edge-on sources due to their intrinsic thickness and bulge
component. This may cause these sources to be shifted towards lower values of
inclination or to be absent from the sample. Indeed there is a hint that at
high inclinations the sample may be slightly biased against low mass galaxies
and that dust-rich spirals in general may appear very red at these inclinations
leading to a possible mis-classification as ellipticals and a bias against
edge-on systems. Nevertheless we believe that our results are
 not affected by strong, inclination-dependent, selection effects, as the
 samples for both ranges of  $\mu_{*}$ are essentially flat in $1 -
 \mathrm{cos}(i_d)$. Furthermore, the distribution of K-band absolute
 magnitudes $M_K$ (which are almost free of of dust attenuation) show no
 inclination dependence
indicative of the presence of a strong selection bias.\newline

Both sub-samples defined in $\mu_{*}$ display considerable scatter in $M_{NUV}$ (after correction for attenuation), with the average inter-quartile range being 1.6, respectively 1.7 magnitudes. This scatter is much larger than can be accounted for by the scatter in $\tau^{f}_{B}$ shown in Fig.~\ref{Fig_corr}. The range of scatter in $\mathrm{M}_{\mathrm{NUV}}$ attributable to the scatter in $\tau^{f}_{B}$, approximated by $\Delta_{s,0.5} \mathrm{log}(\tau^{f}_{B})$ as quoted in \S \ref{tauvmucorr} can only account for a
range of 0.68 respectively 0.9 magnitudes for the ranges $7.8 < \mathrm{log}(\mu_{*}) < 8.3$ and $8.3 < \mathrm{log}(\mu_{*}) < 8.8$ respectively, even in the edge-on case.
Additionally the inter-quartile ranges do not display inclination-dependence, as would be expected if the scatter were predominantly due to object-by-object variations in the dust opacity. Thus, the sample scatter appears to be dominated by the intrinsic scatter in $\mathrm{M}_{\mathrm{NUV}}$.
The histograms of $M_{NUV}$ in Fig. \ref{FIGtest} clearly show that the corrected sample is more peaked and symmetrical with respect to the uncorrected sample, and that the large
shoulder at fainter NUV magnitudes, a product of the inclination dependence, is largely removed after correction. This is the case both for the
optically- and FIR-selected samples, while the remaining breadth of the distribution (especially for the FIR sample) reinforces the conclusion that the scatter in
$\mathrm{M}_{\mathrm{NUV}}$ is intrinsic.
The remaining tail extending to faint NUV magnitudes can most likely be attributed to passive spirals, as presented e.g. in \citet{ROWLANDS2011}, and to contamination caused by early-type galaxies ($\approx$ 5-10\%).\newline 

Overall, we conclude that the inclination-dependent dimming of UV emission from galaxies in the complete optical sample can indeed be predicted using the relation between $\mu_{*}$ and $\tau^f_B$ calibrated on the subset of sources detected in the FIR. The consistency of the correlation with the PT11 model also lends confidence to the supposition that the considerable shift in median magnitude due to the inclination independent component of the attenuation ($\approx 0.9$ and $\approx 1.8$ magnitudes, as predicted for galaxies seen face-on in the ranges of $7.8 < \mathrm{log}(\mu_{*}) < 8.3$ and $8.3 < \mathrm{log}(\mu_{*}) < 8.8$ respectively) is also correct, as this is self-consistently derived together with the inclination-dependent component. This is investigated further in \S~\ref{scatterpsiSM}.\newline

\subsection{The scatter in the specific star formation rate vs. stellar mass relation}\label{scatterpsiSM}
Although we have shown that the $\tau^f_B$ - $\mu_*$ relation in combination with PT11 is effective at predicting the inclination-dependent component of attenuation it is still important to gain a quantitative measure of the efficacy of this technique in predicting the face-on component of the attenuation, which is not so directly probed by the analysis of the attenuation-inclination relation in \S~\ref{attincrel}.
Here we seek to achieve this by utilizing a fundamental
scaling relation between physical quantities derived from
UV/optical emission properties of galaxies
where the intrinsic scatter between the physical quantities
is sufficiently small as to be
exceeded by the scatter in the observed quantities
induced by dust attenuation.\newline

A particularly convenient scaling relation for this analysis is
the well-known relation between specific star-formation rate, $\psi_{*}$,
and stellar mass, $M_{*}$, since, when derived from NUV magnitudes, the
values of SFR used to construct $\psi_{*}$ will be strongly dependent
on the efficacy of the method used to correct for attenuation,
whereas, as shown by \citet{TAYLOR2011} and discussed in \S~\ref{correlation}, the
values of $M_{*}$ are much less affected by dust. Here we convert from de-attenuated values of $M_{NUV}$ to $\psi_{*}$ using the conversion given in \citet{KENNICUTT1998} scaled from a \citet{SALPETER1955} IMF to a \citet{CHABRIER2003} IMF as in \citet{TREYER2007}. We note that the exact choice of conversion factor has no bearing on the analysis.
\newline

\begin{figure*}
\plotone{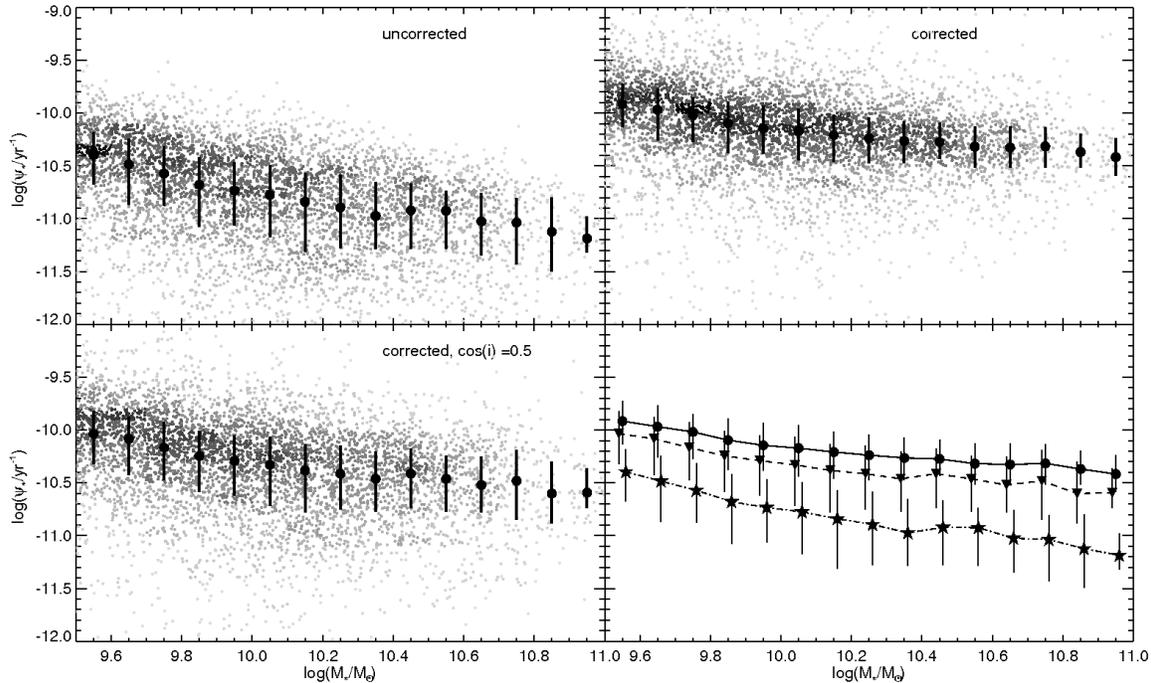}
\caption{Specific star formation rate $\psi_{*}$ as a function of stellar mass
  $M_{*}$ for a subsample of the \textit{OPTICAL} sample with $7.6 \le
  \mathrm{log}(\mu_{*}) \le 9.0$ and $M_{*} > 10^{9.5}\,M_{\odot}$. The
relation is shown before correction for attenuation by dust (top left panel),
after the full inclination-dependent correction, described in \S~\ref{deriveatcor},
using the PT11 model in conjunction with the $\tau^f_B\,-\,\mu_{*}$
relation (top right panel), and after a partial correction using the procedure
of \S~\ref{deriveatcor} but artificially setting a 
uniform inclination $i_d$ with $\mathrm{cos}(i_d) = 0.5$ for all galaxies
(lower left panel). The sources are binned in 15 bins of equal size in $M_{*}$, with the median depicted by a filled circle, and the bars showing the interquartile range. The scatter in the relation due to the scatter in the NUV is reduced from the uncorrected to the fully corrected case. The intrinsic values of $\psi_{*}$ are shifted upwards w.r.t. the uncorrected values. 
The linear gray-scale shows the number density of sources at that position, with the same scale having been applied to all samples.
The median values and interquartile ranges are shown together in the bottom right panel. The uncorrected values are depicted by stars and a dash-dotted line, the values corrected at a fixed inclination of $\mathrm{cos(i)}=0.5$ are shown as inverted triangles and a dashed line, and the fully corrected values are shown as circles and a solid line. The bin centers have been offset by 0.01 in $\mathrm{log}(M_*)$ for improved legibility.} 
\label{Fig_SSFR}
\end{figure*}

In Fig.~\ref{Fig_SSFR} the $\psi_{*}$ vs. $M_{*}$ relation is
plotted for the \textit{OPTICAL} sample limited to
$7.6 \le \mathrm{log}(\mu_{*}) \le 9.0$ and $M_{*} > 10^{9.5}\,M_{\odot}$
(following the range of applicability of the $\tau^f_B\,-\,\mu_{*}$ 
relation
given in \S~\ref{correlation}). To differentiate between the effects of the 
corrections
for the face-on and inclination-dependent components of attenuation we 
plot
the relation three times: with no attenuation corrections (top left);
with attenuation corrections as detailed in \S~\ref{deriveatcor}
, but
with all inclinations artificially set to the median inclination of the
sample (bottom left); and with the corresponding full 
inclination-dependent
corrections (top right).
The expected trend of decreasing $\psi_{*}$ as a function of $M_{*}$
is seen in all three panels. Comparison of the top left and top right panels
shows, however, that the application of the inclination dependent
attenuation corrections, in addition to inducing a overall
systematic shift by a factor of $0.6\,$dex
in $\psi_{*}$, reduces
the scatter in the relation, from $0.63\,$dex in the uncorrected
relation to  $0.43\,$ dex in the corrected relation\footnote{All measurements of scatter were calculated
as the difference between
the quartiles of the distribution in $\psi_{*}$, averaged over
15 equal sized bins in $M_*$, and weighted by the number of galaxies in each bin.}\newline

This suggests
a substantial predictive power, both of the $\tau^f_B\,-\,\mu_{*}$ 
relation
and the PT11 model, since we have applied an object-specific
and large multiplicative correction to the NUV fluxes (by
factors ranging from 2.5 to 6.3 interquartile with a median correction of 3.8), yet
have nevertheless succeeded in markedly reducing the logarithmic scatter in the
$\psi_{*}$ vs. $M_{*}$ relation\footnote{ As shown from the analysis of the multivariate relation between
$\tau^f_B$, $M_{*}$, and $\mu_{*}$ in \S~\ref{correlation} the spread in face-on optical
depth at a given $M_{*}$ arises from the large spread in disk radii
for galaxies of a given $M_{*}$, in conjunction with the
close-to-linear $M_d$ vs. $M_{*}$ correlation}.
Furthermore, comparison of the scatter in
the partially corrected relation in the bottom left panel
($0.57\,$dex) with the $0.63\,$dex
scatter in the uncorrected relation in the top left panel
shows that the total reduction in scatter is due not only to the 
correction
of the inclination-dependent component of the correction, but also due to
the correction of the face-on component of the correction.
This is a strong indication that the zero point of the
attenuation corrections (i.e., the face-on attenuation predicted
by the PT11 model, which is the major contributor to the
total attenuation) cannot be strongly in error.
If the face-on component of the attenuation would have been
independent of the stellar mass surface density,
the large range of predicted face-on optical depths at a fixed
stellar mass would have increased the scatter, rather than have
diminished it.\newline

We note that the intrinsic scatter of $0.43\,$dex (interquartile)
in the corrected relation of Fig.~\ref{Fig_SSFR} (top right panel) will have substantial contributions
from random errors.
Major sources of this random uncertainty probably arise from measurement uncertainties in the determination of disk surface areas as well as from the estimates of galaxy stellar mass ($\sim 0.1\,$dex).
In addition, we recall that the galaxy sample will be contaminated at the 5-10 \% level by mis-classified spheroids. Furthermore we cannot rule out that there is some significant intrinsic scatter in the $\tau^f_B\,-\,\mu_{*}$ 
relation which would also induce a component of scatter in the corrected $\psi_*$ - $M_*$ relation.
All this suggests that the intrinsic scatter in the $\psi_{*}$ vs. $M_{*}$ relation for spiral galaxies may be very low.

\subsection{Implications for the distribution and optical properties of grains
  in galaxy disks}\label{structimplications}  

The success of the $\tau^f_B\,-\,\mu_{*}$ relation in combination with the PT11 model in predicting both the face-on and inclination dependent component of the attenuation in
spiral galaxies has implications both for the spatial distribution of grains in galaxian disks as well as for the optical properties of these grains.\newline 
   
Firstly, the quantitative consistency
between the measured dust surface density and the inclination-dependent
attenuation of stellar light in disk galaxies, as predicted
by PT11, is consistent with most of the dust in disks being distributed
in structures sufficiently large to have a substantial projected surface
filling factor.
Furthermore, recalling that the $\tau^f_B\,-\,\mu_{*}$ relation is calibrated
using measurements of the total submm flux, i.e tracing the total mass of dust in galaxies,
the reduction in scatter about the $\psi_{*}$ - $M_{*}$ relation induced by the application of PT11
points qualitatively towards most of the mass of dust in spiral galaxies (as traced in the submm) being distributed in
diffuse, translucent structures, with a large fraction of the grains
being exposed to UV in the diffuse interstellar radiation field as assumed by the PT11 model.\newline

\begin{figure}
\plotone{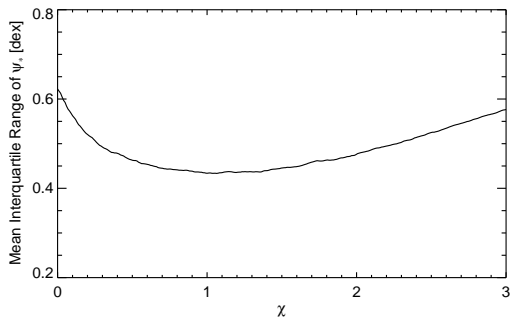}
\caption{Weighted mean interquartile range of $\psi_{*}$ as a function of $M_{*}$ derived for fractions $\chi \cdot \tau^f_{B}$ of $\tau^f_{B}$ sampled in steps of 0.01. The minimum value of 0.43 is attained for $\chi = 1.06$, however, it is not significantly distinguishable from that of  $\chi = 0.95, \dots, 1.1$.  }
\label{Fig_IQR}
\end{figure}

In order to make this statement more quantitative we have in Fig.~\ref{Fig_IQR} plotted the mean interquartile range in the $\psi_*$ - $M_*$ relation as a function of attenuation corrections based on an effective value of the dust opacity parameterized by a multiplicative scalar value $\chi$ as $\chi \cdot \tau^f_{B}$.  
If, contrary to the model of PT11 in which $\gtrsim 85$\% of the total dust mass is diffusely distributed, a large fraction of the dust mass measured in the submm were contained in compact, highly self-shielded regions, not exposed to the diffuse interstellar UV radiation field,
the minimum in scatter about the $\psi_{*}$ - $M_{*}$ relation should be attained for a relatively small value of $\chi$ (i.e. $ \chi << 1$).
Instead we find that the minimum scatter is attained for $\chi = 1.06$, but that a range of 
$\chi \approx 0.95 - 1.1$ is not significantly distinguishable. This implies that, consistent with the PT11 model a fraction of $\gtrsim85$\% of the total dust mass is distributed in diffuse, translucent structures. 
A more detailed analysis
of the dependence of scatter in scaling relations, though beyond the
scope of this paper,
could in principle be used to fine tune model assumptions
about the fraction of dust in clumps which are heavily self-shielded to UV light
in disks of spiral galaxies, and thereby further improve
estimates of the absolute level of $\psi_{*}$ in the
relation as well as the intrinsic scatter of the physical
quantities.\newline

An analysis of the type performed here also has the potential to empirically constrain the ratio between grain emissivities in the submm and UV/optical range. 
This arises because, while the estimates of $\tau^f_b$ on which the $\tau^f_B\,-\,\mu_{*}$ relation is based are directly proportional to the dust emission coefficient in the FIR/submm, the amplitude of the attenuation corrections depends upon the dust
emission coefficient at (in this case) UV wavelengths. Specifically, the
demonstrated ability to correct for the inclination-dependent and face-on components of
attenuation using
Eq.~\ref{correleq}, which was derived and calibrated using the FIR, is
consistent with the ratio of the UV/optical and submm grain absorption coefficients being as
described in the model of \citet{WEINGARTNER2001}. \newline

\section{Discussion}
\label{Discussion}
\subsection{Dust production in spiral galaxies}
As already noted in \S~\ref{correlation}, the almost linear relation between
the opacity of a galaxy disk, $\tau^f_B$, and the surface density
of stellar mass, $\mu_{*}$, appears to arise mainly from a
roughly linear relation between the extrinsic quantities
of total stellar mass, $M_{*}$, and total dust mass, $M_{d}$.
Although only established directly for galaxies with available
FIR/submm measurements, the applicability of the same $M_{*}\,-\,M_{d}$
relation to a statistically much more complete population of optically
selected spiral galaxies from the GAMA survey is consistent with our
analysis of the attenuation-inclination relation and $\psi_*$ - $M_*$ relation presented in \S~\ref{test}.\newline

The physical origin of this link between $M_{*}$ and $M_{d}$
is far from obvious. On the one hand, the stellar mass is dominated by
old, low mass stars which formed early on in the ca. 10 Gyr lifetime of
a typical spiral galaxy. On the other hand, the main known process of
injection of dust grains into the ISM is the condensation of metals
in the atmospheres of AGB stars on timescales of
$\sim 2 \pm 1\,\cdot 10^9\, \mathrm{yr}$ \citep{DWEK1980,MCKEE1989,MORGAN2003,
FERRAROTTI2006,ZHUKOVSKA2008,GAIL2009,JONES2011}
\footnote{Refractory grains have been observed to form in
the metal-rich ejecta of core-collapse supernovae, which, alongside
type~Ia supernovae, have also been postulated to be major sources of
interstellar grains.  However, with the possible exception of the remnant of SN1987A \citep{MATSUURA2011,LAKICEVIC2012}, FIR/submm measurements of cold unshocked
ejecta in the central regions of prototypical young supernova remnants
(SNRs) have shown that the ratio of solid state to gas phase ejecta
is modest in comparison to the grain-to-gas ratio in the ISM
\citep[e.g.,][]{TUFFS1997,GREEN2004,BARLOW2010,GOMEZ2012}.
Given that, to escape the SNR, the condensates must
traverse the very shocks postulated to be the main sink for
refractory grains in the ISM, it seems unlikely that supernovae are major
primary sources of interstellar grains.
}
much shorter than the ages of spiral galaxies.
Moreover, detailed modelling of the life cycle of refractory grains in the
solar neighborhood \citep[e.g.,][]{JONES1996,JONES2011}
predict that grains in the Milky Way are destroyed by SNe shocks
in the tenuous ISM on timescales of $\sim 10^8\,\mathrm{yr}$,
much shorter than the timescale for the injection of dust
from AGB stars, requiring that almost all observed
refractory dust in the diffuse ISM must have been (re-)formed in situ
soon after its destruction. This picture of rapid destruction and
formation in the ISM is, however, difficult to reconcile with key
physical and chemical properties of pre-solar grains as found in 
meteorites,
most notably the segregation into separate populations of silicate and
carbonaceous grains with a high abundance of minerals similar to
those known to be produced in stellar sources. As discussed in detail
by \citet[]{JONES2011}, one is consequently confronted with a conundrum:
either the grain destruction rates in the ISM have been grossly
overestimated, allowing most refractory grains in the ISM to have an 
origin in
AGB stars, or, alternatively, an as yet unidentified but very efficient
mechanism exists that can convert gaseous metals in a low temperature and
low pressure ISM into solid particles with the observed physical and 
chemical
characteristics of interstellar grains.\newline

The close to linear relation between $M_{d}$ and $M_{*}$ underlying the
$\tau^f_B\,-\,\mu_{*}$ relation naturally favors the
existence of a mechanism for efficient growth of
refractory dust out of the ISM, since any such mechanism would
tap into the full reservoir of metals in the ISM, which are
related to the integrated star formation over the lifetime of a galaxy.
By contrast, if interstellar dust were mainly composed of longer-lived
grains injected by AGB stars on timescales of
$\sim 2 \pm 1\,\cdot 10^9\, \mathrm{yr}$, one would expect the total mass
of dust to be approximately proportional to the star formation rate (SFR)
multiplied by a residency time, at least for systems with ages larger than the average dust destruction timescale\footnote{In the very early evolution of systems dust mass may increase in parallel with stellar mass simply due to continuos injection of dust into the ISM driven by star formation. Only after the the age of the system increases to more than the average dust lifetime can the mass of dust be expected to be proportional to the SFR multiplied by a residency time.}. Since the residency time should decrease
with increasing SFR (since the frequency of destructive SN shocks
should be proportional to the SFR), and the SFR per unit stellar mass
is known to decrease as a function of stellar mass, a
strongly sublinear dependence between $M_{d}$ and $M_{*}$ would be
predicted, even if an increase in the dust yield as a function of
metallicity is taken into account. As such, if the origin of dust grains in the ISM 
were predominantly stellar, one would expect an at most very weak dependence of
dust mass on stellar mass.\newline

If the mechanism for growth of grains out of the ISM
implied by the slope of
the $\tau^f_B$ - $\mu_{*}$ relation was sufficiently prompt and efficient
it would ubiquitously lead to a high fraction (i.e, of the order unity),
$\eta$ of all refractory elements being condensed into grains in the
ISM of all spiral galaxies 
(as also inferred by,  e.g.  \citealt[][]{DWEK1998,EDMUNDS2001}, and
\citealt[][]{DRAINE2009}). 
To test whether $\eta$ really does assume   
a universally high value in the ISM of local Universe spiral
galaxies, we can make use of the well-established empirical relations
linking stellar mass with gas phase metallicity and
gas mass for this galaxy population.
Specifically, the product of these relations will yield
a relation between total metal mass and $M_{*}$, which, by multiplying
the metal mass by a constant value for $\eta$, will predict a
relation between total
dust mass $M_{d}$  and $M_{*}$.\footnote{The underlying assumption of such a prediction is that all galaxies will have experienced a similar star formation history. Variations in this history can give rise to significant scatter around the relation, in particular, the time at which a large burst of SF occurs may strongly influence the observed dust-to-stellar mass ratio. Nevertheless, as we have endeavored to select a pure sample of normal spiral galaxies, and the specific SFRs obtained for the \textit{OPTICAL+FIR} sample don't display bimodality, such an assumption does not appear unreasonable.}. This relation
can then be compared with the observed relation.\newline

\begin{figure}
\plotone{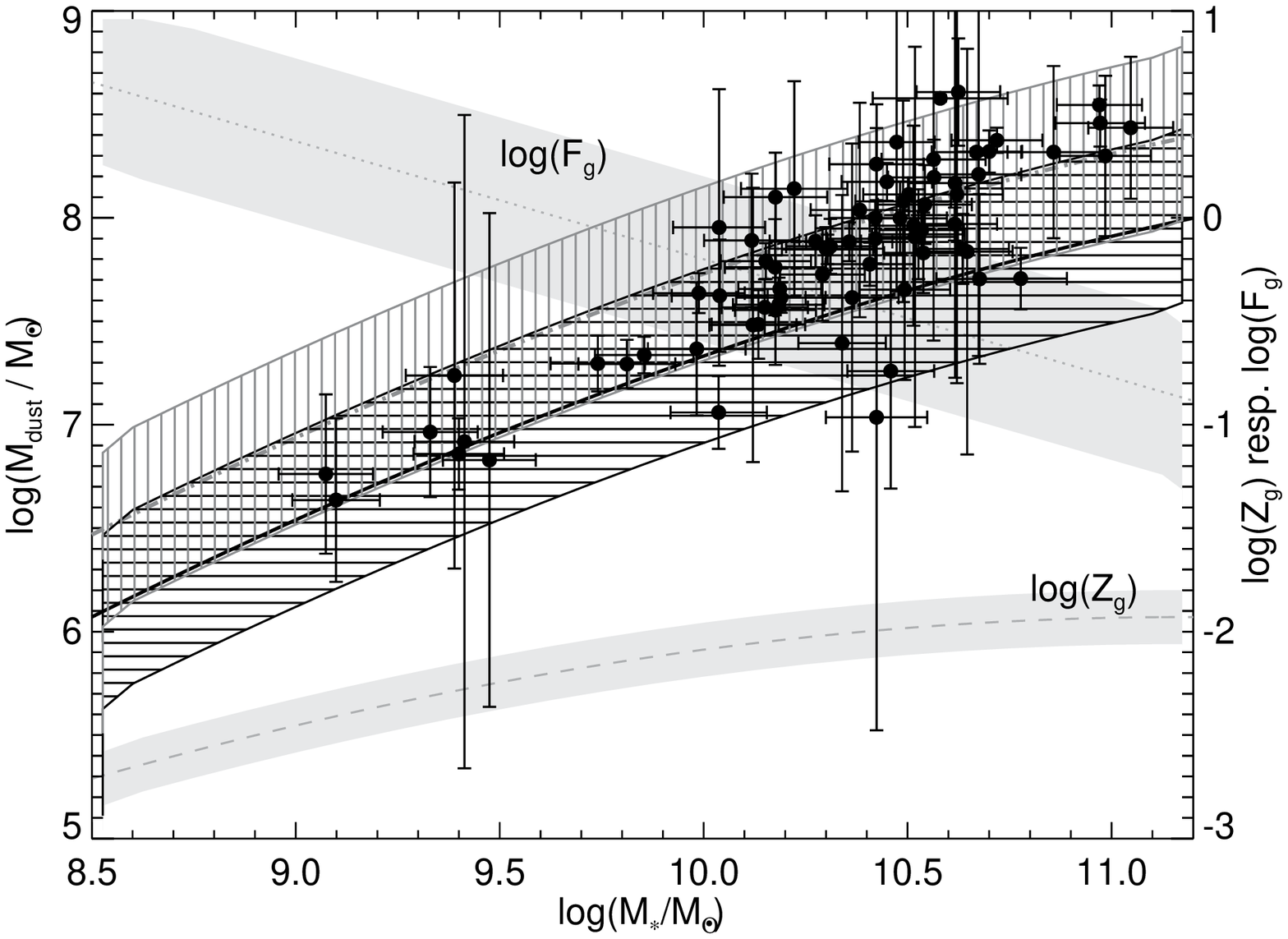}
\caption{Predicted values of dust mass $M_{\mathrm{dust}}$ as a function of
  stellar mass $M_{*}$ for an assumed conversion of a fraction $\eta$ of all
  ISM metals to dust. The mass-metallicity relation
  \citep{TREMONTI2004,KEWLEY2008} converted to gas-phase metallicities, a \citet{CHABRIER2003} IMF as in PS11 and the stellar-to-gas mass ratio
(PS11) used in deriving the expectations are overplotted as a dashed 
and dotted lines, respectively with the shaded areas indicating the range of 1-$\sigma$ scatter around the relations. 
The predicted relation and 1-$\sigma$ scatter (derived as sum quadrature) between $M_{\mathrm{dust}}$ and $M_{*}$ is shown for $\eta=0.5$ ( solid black line and horizontally striped region) and for $\eta=1.$ (dash-dotted gray line and vertically striped region). The diffuse dust masses of the \textit{OPTICAL+FIR} sample, derived from the values of $\tau^f_B$ using Eq.~\ref{taubf1} are overplotted as filled circles with error bars (errors on $M_{\mathrm{dust}}$ take into account errors on $\tau^f_B$ and $\theta_{e,ss,r}$)}. 
\label{Fig_pred}
\end{figure}

Fig.~\ref{Fig_pred} shows the predicted relations between $M_{*}$
and $M_{\mathrm{dust}}$ as derived using the mass-metalicity
relation for late-type galaxies \citep{TREMONTI2004,KEWLEY2008}, converted to
gas-phase metalicities and a \citet{CHABRIER2003} initial mass function (IMF)
as in \citet[][PS11]{PEEPLES2011}, and the gas-to-stellar mass ratio from 
PS11. 
The relations for $\eta=0.5$ and $\eta=1$ are respectively shown by the
solid black and dash-dotted gray lines, together with horizontally and
vertically striped regions indicating the 1-$\sigma$ scatter around the
relations. It is apparent that the observed trend in
$M_{*}$ vs. $M_{\mathrm{dust}}$, shown by the overplotted data points from
the \textit{OPTICAL+FIR} galaxies, is indeed quite well predicted 
by the mass-metalicity and gas mass vs. stellar mass relations for
constant $\eta$, and that the required value of $\eta$ indeed has to be
high. If fact values of between 0.5 and 1 are required for $\eta$,
about a factor of two higher than the several tens of percent of
ISM metals that are predicted to
be present in the form of grains by several detailed
physical models such as those by \citet[][$\eta\sim0.4$]{DWEK1998},
\citet[][$\eta\gtrsim0.4$]{EDMUNDS2001}, and
\citet[][$\eta\sim0.3$]{GALLIANO2008}. However, as we discuss in Appendix~\ref{AppendixTAUBF2},
it must be born in
mind that the measured dust masses plotted in Fig.~\ref{Fig_pred}
were derived from the FIR/submm observations using a mass absorption
coefficient which is actually quite uncertain. 
In particular,
whereas the relative values of the UV and submm
grain absorption cross sections of \citet{WEINGARTNER2001}
used by our radiation transfer analysis
connecting the submm emission and UV attenuation characteristics of
spiral galaxies have been empirically constrained with respect to hydrogen
gas column through measurements of extinction and emission of diffuse dust
in the Milky Way, the absolute value of the absorption cross section per unit
grain mass $\kappa_{\mathrm{m}}$, needed to deduce the value of $\eta$, is relatively uncertain.
As noted by \citet{DRAINE2007} the value of $\kappa_{\mathrm{m}}$
for the model of \citet{WEINGARTNER2001}
requires more heavy elements than appear to be available and the
mass of dust may be overestimated by a factor of $\sim1.4$.
\footnote{This is also reflected by the fact that determinations
of $\kappa_{\mathrm{m}}$
based on metal abundance as an {\it input constraint}
predict higher submm grain
emissivities than those in the \citet{WEINGARTNER2001} dust model.
For example the emissivity model used by \citet{DUNNE2011} is
partly based on the analysis of \citet{JAMES2002}, who,
by assuming that $45.6$\% of all metals are converted to dust,
derived a dust absorption coefficient per unit mass at a wavelength of
850${\mu}m$ which is  $\sim70$\% larger than
that of \citet{WEINGARTNER2001}.}
Such a shift in $\kappa_{\mathrm{m}}$ (while leaving the grain
absorption cross sections relative to hydrogen unchanged)
would reconcile the majority of the measurements
plotted in Fig.~\ref{Fig_pred} to an $\eta$ of $\sim$0.5, given the observed
scatter. At the same time this would preserve the observed quantitative
connection between the UV attenuation
and the observed surface density of
submm emission, as predicted by the PT11 radiation transfer model.\newline

The only way we could envisage maintaining this demonstrated ability to link
attenuation of starlight to the observed surface density of submm emission,
while avoiding $\eta$ approaching unity {\it and} avoiding having to raise
$\kappa_{m}$ for dust in the diffuse ISM from the values given by
\citet{WEINGARTNER2001}, would be to invoke a population of highly
self-shielded compact dense clumps as the source for a large fraction of the observed
submm emission from spiral galaxies, in conjunction with us having
systematically overestimated the intrinsic angular sizes of the disks seen in
r-band.\footnote{ This would reduce the mass of dust needed to explain the observed submm
fluxes, since the dust in self-shielded clumps, while no longer able to efficiently
participate in the attenuation of light from stellar populations not spatially
correlated with the clumps, might be expected to have a much higher value of
$\kappa_{\mathrm{m}}$ in the submm, due to the formation of ice mantles in such environments.
The corresponding reduction in the mass of dust in the diffuse ISM would then need to
be exactly compensated for by the reduction in the inferred intrinsic angular size of
the disk, such as to restore the opacity of the disk to the levels needed to predict
the attenuation of the starlight (as quantified through the attenuation-inclination
relation and the scatter in the $\psi\,-\,M_*$ relation).}  
While the present accuracy 
of measurement of intrinsic disk sizes, as outlined in Appendix~B, may not completely 
rule out such a scenario, recent high angular resolution
submm imaging of the galactic plane of the Milky Way by the Herschel Space Observatory, 
sensitive to emission on all angular scales, clearly show that the vast majority of 
submm photons originate from translucent large scale structures
\citep[e.g.][]{MOLINARI2010}.\newline

We conclude that  the near linearity and high constant of
proportionality of the $\tau^f_B\,-\,\mu_{*}$ relation,
itself based on a near linear relation between $M_{\mathrm{dust}}$ and $M_{*}$,
is indeed in good agreement with a roughly constant and high (but still physical)
fraction of all ISM metals being present in the form of grains, and is
best understandable in terms of the existence of a
ubiquitous and very rapid mechanism for the in situ growth of grains in
the gaseous ISM. Based on a joint consideration of
measured dependencies of dust mass, gas fraction and metallicity
on stellar mass, our simple analysis provides a direct and 
model-independent
empirical confirmation of work which has used more sophisticated chemo-dynamical simulations
of the dust cycle in local and high-z galaxies applied to dust abundances and gradients
to infer a dominant in situ source of interstellar grains both in local, normal galaxies and high-z starbursts 
\citep[e.g.][]{DWEK1998, CALURA2008, DRAINE2009, MICHALOWSKI2010, DUNNE2011, DWEK2011a,DWEK2011b, INOUE2011,MATTSON2012,VALIANTE2011}.
Since we make no assumptions about the stellar populations other
than that of the current injection rate of stardust being proportional
to the recent SF rate, our conclusion that stardust is a minor constituent
of dust in spiral galaxies holds even if the initial mass function for
stars were to be top heavy, which has been suggested 
\citep[e.g. by][]{DUNNE2011} as a possible
way of alleviating the need for grain growth in the ISM.\newline

Moreover, the ability of the $\tau^f_B$ - $\mu_{*}$ relation to
predict the NUV attenuation-inclination relation suggests that
the majority of grains are exposed to non-ionizing UV light in the
diffuse interstellar radiation field, so are refractory in nature\footnote{Volatiles in the form of ices will almost instantaneously
return to the gas phase through photodesorption if exposed to
UV in the diffuse interstellar radiation field} and reside in the diffuse ISM.
As a consequence, our results not only require a very efficient
grain formation mechanism, but also require that this mechanism
pertains to the formation of refractory grains, rather than merely to the
condensation of ices in highly self-shielded regions.\newline

The nature of the mechanism for forming refractory grains in the ISM
is completely open.  In their comprehensive analysis of the evolution
of the interstellar dust population in the solar neighborhood in the Milky 
Way,
based on a one zone chemical evolution model accounting for the growth of individual species, \citet{ZHUKOVSKA2008} conclude that 
the
interstellar dust population is dominated by refractory grains grown
by accretion of gas phase metals in dense molecular clouds, with stardust
(in their model from both from AGB stars and from supernovae)
constituting only a minor fraction.
This result is consistent with our conclusions,
independently inferred from the $\tau^f_B\,-\,\mu_{*}$ relation for spiral
galaxies, but only provided a mechanism exists to propagate the refractory
grains from the clouds into the diffuse ISM on timescales shorter than
the timescale for grain destruction in the diffuse ISM.
Alternatively, \citet{DRAINE2009} has proposed that refractory grains
can grow in diffuse interstellar clouds, in the presence of
UV radiation. This would seem to be more easy to reconcile with our
result that the bulk of all grains must reside
in translucent structures illuminated by UV, as it would bypass the need for a rapid propagation mechanism.\newline

In general, the $\tau^f_B\,-\,\mu_{*}$ relation may be useful
as a diagnostic tool to investigate the universality
and nature of the in situ grain-formation mechanism.
In particular, although our analysis favors grain condensation
from the ISM as the main grain injection mechanism,
our present statistics cannot rule out that a significant minority of
the grains have a stellar origin. \citet{JONES2011} emphasize that there is a considerable uncertainty
in theoretical predictions for grain lifetimes, so that, while
there is a strong requirement for Carbonaceous grains to be
rapidly recycled in the ISM, this requirement may be less strong
for Silicate grains. Analysis of the $\tau^f_B\,-\,\mu_{*}$ relation
for larger statistical samples will allow separate relations
to be established for spiral galaxies as a function of recent SF history,
spiral arm coverage (i.e., lateness) and specific SF rate which
may throw more light on this question, particularly if accompanied with
data on the strength of the $2200\,\mathring{\mathrm{A}}$ absorption and MIR Polycyclic Aromatic
Hydrocarbon (PAH) emission features, both of which specifically
probe Carbonaceous particles.

\subsection{The attenuation of starlight in spiral galaxies}
\begin{figure*}
\plotone{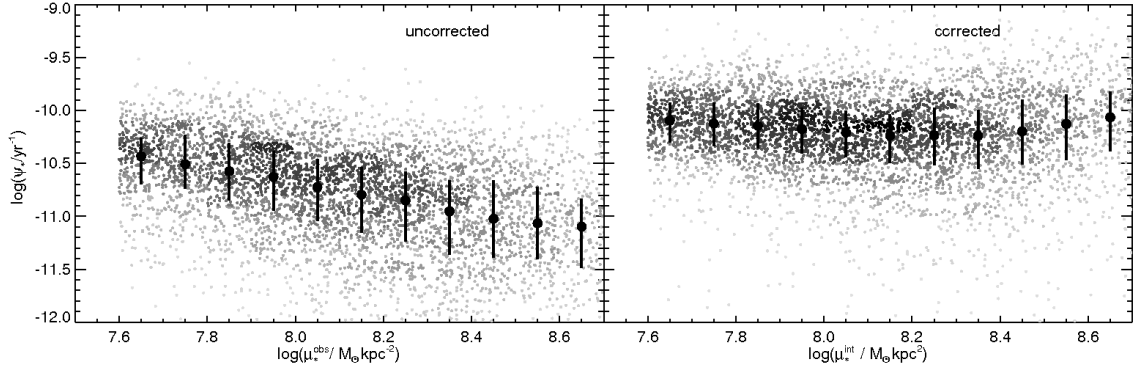}
\caption{Specific star formation rate $\psi_{*}$ as a function of stellar mass
  surface density $\mu_{*}$ for a subsample of the \textit{OPTICAL} sample with
  $7.8 \le \mathrm{log}(\mu_{*}) \le 9.$ and $M_{*} >
  10^{9.5}\,M_{\odot}$. The left panel shows uncorrected values of $\psi_{*}$
  , while the right shows the corrected values of $\psi_{*}$.
Here again, the scatter is reduced and, notably, the slope of the relation is altered w.r.t. the uncorrected quantities. The median values of $\psi_{*}$ for bins of equal size in $\mu_{*}$ are shown as filled circles, with the bars depicting the interquartile range. The notable increase in scatter at high values of $\mu_{*}$ as well as the increase in $\psi_{*}$ may be caused by contamination from nuclear starbursts. The linear gray-scale shows the number density of sources at that position, with the same scale having been applied in both panels. }
\label{Fig_SSFRmu}
\end{figure*} 
Having discussed the use of the $\tau^f_B\,-\,\mu_{*}$ relation as a diagnostic of
physical processes driving the efficient
production of interstellar dust in the disks of spiral galaxies, we return to the main goal of this investigation, namely
the use of the $\tau^f_B\,-\,\mu_{*}$ relation, in conjunction with the
radiation transfer model of PT11, to correct for the attenuation of
stellar light by dust.
As described in \S~\ref{test},
this can be done on a object-to-object
basis for large statistical samples of spiral galaxies, using readily available optical
photometric properties for each galaxy. Although the predicted attenuations are quite substantial, especially in the UV, our quantitative 
analysis of the attenuation-inclination and of the scatter in the $\psi_*$ - $M_*$ relation lends some confidence that the corrections are
not largely in error.\newline
 
As we have already noted in the case of the $\psi_*$ - $M_*$ relation the scatter in fundamental scaling relations based on UV and optical quantities can be 
significantly reduced through application of attenuation corrections based on the $\tau^f_B\,-\,\mu_{*}$ relation even when no
dust emission data is available.
The $0.43\,$dex interquartile scatter in the $\psi_*$ - $M_*$ relation after correcting for attenuation already implies a 
very tight relation between the current and past
star-formation in spiral galaxies in the local Universe
that would need to be reproduced by any theory of the formation
and growth of spiral galaxies.\newline

Although affecting the scatter of the $\psi_{*}$ vs. $M_{*}$ relation,
the attenuation corrections do not, as already noted, strongly affect the
slope of this relation, at least in the range of $M_*$
for which the attenuation corrections are presently available through
the $\tau^f_B\,-\,\mu_{*}$ relation. This is
because, as shown in \S~\ref{correlation}, opacities are statistically
much more tightly related to stellar mass surface density,
rather than to stellar mass, coupled with the fact that spiral
galaxies of a given stellar mass exhibit a wide range of disk sizes.
However, this situation will no longer apply to scaling relations
as a function of the stellar surface mass density, $\mu_{*}$.
To illustrate this, we plot in
Fig.~\ref{Fig_SSFRmu} the relation between $\psi_{*}$ and $\mu_{*}$, for the same
sample as used for the $\psi_{*}$ vs. $M_{*}$ relation, both
before and after correction. Remarkably, the slope in the relation
between the uncorrected quantities is entirely removed after correction
for dust, showing that $\psi_{*}$ is statistically independent of
$\mu_{*}$. The scatter in the relation is reduced from $0.62$ to $0.49\,$dex.
The latter value is somewhat larger than that for the $\psi_{*}$ vs. 
$M_{*}$ relation,
perhaps implying that the $\psi_{*}$ vs. $M_{*}$ relation has a smaller
intrinsic scatter and is thus the more fundamental relation.\newline

Overall, the fact that disk opacities scale
systematically with stellar mass surface density,
as opposed to being randomly distributed,
may help to explain the preservation of
systematic and in some cases surprisingly tight relations
between optical or UV tracers of physical quantities,
even when these observational tracers
are heavily affected by dust attenuation, and may help to explain why
many relations were historically discovered only with relatively crude
corrections for dust attenuation. Apart from the scaling relations
analyzed here, a further relation which would be particularly pertinent to
reanalyze would be the Tully-Fisher (TF) relation between luminosity
and dynamical mass \citep{TULLY1977}, which is
even tighter than the $\psi_{*}$ vs. $M_{*}$ relation. A  similar analysis
to that applied to the $\psi_{*}$ vs. $M_{*}$ relation here applied to
the TF relation could provide a still sharper tool for statistical
analysis of attenuation corrections, as well as potentially improving
the accuracy of the TF relation at shorter wavelengths, both as a distance
indicator, and as a diagnostic of the formation and evolution
of spiral galaxies.\newline

Finally, we re-emphasize that the corrected relations are only for subsets
of the galaxy population restricted in $M_{*}$ and $\mu_{*}$
according to the limits of our present calibration of the
$\tau^f_B\,-\,\mu_{*}$ relation as defined in \S~\ref{correlation}.
In order to establish attenuation-corrected scaling relations for the 
whole
population of disk galaxies, it is crucial
to extend knowledge of FIR/submm emission, and thereby calibration
of the $M_{*}$ and $\mu_{*}$ relation, to a representative subset of
the entire population of rotationally supported disk galaxies,
including dwarf galaxies.\newline

\section{Summary and outlook}
\label{summary}
We have presented a correlation between the face-on B-band dust-opacity $\tau^{f}_{B}$ and the stellar mass surface density
$\mu_{*}$ for normal late-type galaxies for a range in $\mu_{*}$ (in units of
$M_{\odot}kpc^{-2}$) of $7.6 < \mathrm{log}(\mu_{*}) < 9.0$. Using the attenuation-inclination relation 
for rotationally supported late-type galaxies we have demonstrated that the values of $\tau^{f}_{B}$ estimated by means of this correlation successfully
correct statistical samples of late-type galaxies for dust attenuation.
In order
to apply the correlation to complete samples of spiral galaxies
in terms of $M_{*}$, the present depth of FIR/submm data used to
derive the $\tau^{f}_{B}$ - $\mu_{*}$ correlation means such samples
need to be limited to $M_{*} \geq 10^{9.5} \mathrm{M}_{\odot}$. While the selection method employed to select spirals depends on the
availability of NUV and optical data, only photometric optical data is needed to estimate the values of $\tau^{f}_{B}$, as the required S\'ersic fits 
are performed on optical (\textit{r}-band) imaging, and stellar masses have
been derived using only optical data. As such this
correlation presents a means to obtain attenuation corrections for samples of galaxies on the basis of optical photometric data alone, and is thus
applicable to a large range of data-sets.\newline

We quantitatively demonstrate the efficacy of attenuation corrections
using the  $\tau^f_B$ - $\mu_{*}$ relation in concert with the PT11
model through analysis of the attenuation - inclination
relation in the NUV and the $\psi_{*}$ vs. $M_{*}$ relation
for a large sample of local Universe spiral galaxies.
Both the inclination-dependent and the face-on components of attenuation
are well predicted; for the latter we find that the scatter in the $\psi_{*}$
vs. $M_{*}$ relation is minimized for corrections within
10 percent of those estimated using the $\tau^f_B$ - $\mu_{*}$ relation and the PT11 model,
consistent with the large majority of dust residing in the diffuse ISM, rather than in opaque clouds.
Overall our results are consistent with a general
picture of spiral galaxies in which most of the submm emission originates 
from grains residing in translucent structures exposed to UV in the
diffuse interstellar radiation field.
\newline

We identify a roughly linear relation between the dust-mass and stellar
mass of a galaxy for the  range of $M_{*}$ probed by our sample to be the
dominant driver for the   $\tau^f_B$ - $\mu_{*}$
correlation. Combining this result with known empirical dependencies of the
gas-to-stellar mass ratio and metallicity on stellar mass, we make a largely
model-independent inference requiring the rapid (re-)formation of dust grains
after their destruction in the ISM, as has also been inferred from
chemo-dynamical modelling of the dust cycle in the Milky-Way and
external galaxies. Our requirement that the reformed dust is refractory and
largely resides in the diffuse ISM (rather than in dense clouds) will
further constrain models of the dust cycle in galaxies.
We posit that the $\tau^f_B$ - $\mu_{*}$ relation applied to statistical samples of late-type galaxies subdivided by, e.g, morphology/lateness and SFR, may provide a diagnostic tool to investigate the applicability to late-type galaxies in general of results pertaining to grain formation derived from studies of the local ISM in the Milky Way.\newline

At the moment, the determination of $\tau^f_B$ using modified black body functions as well as the uncertainty, introduced by the effects of diffuse dust and the bulge-to-disk ratio, on the measured scale lengths of galaxies are major contributors to the uncertainties in the results presented. On the other hand, however, the approach chosen in this paper which gives rise to these uncertainties, also ensures that the result given by Eq.~\ref{correleq} is not strongly model dependent, and is straightforward to implement in practical applications. In the future we will return to this topic using a larger data set and compare these results with such obtained using values of $\tau^f_B$ and $\theta_{s,d,r}$ stemming from  a fully self-consistent radiation transfer modelling approach (cf. PT11).
Future work will also focus on extending the range in $\mu_{*}$ (hence also in $M_{*}$) over which the relation is applicable using fully self-consistent radiation transfer modelling, two component S\'ersic fits (using future higher resolution photometry as the data becomes available from VST KIDS, VISTA VIKING) and a larger data set in the form of the entirety of H-ATLAS data.

\section*{Acknowledgements}
GAMA is a joint European-Australasian project based around a spectroscopic campaign using the Anglo-Australian Telescope. The GAMA input catalogue is based on data taken from the Sloan Digital Sky Survey and the UKIRT Infrared Deep Sky Survey. Complementary imaging of the GAMA regions is being obtained by a number of independent survey programs including GALEX MIS, VST KIDS, VISTA VIKING, WISE, Herschel-ATLAS, GMRT, and ASKAP providing UV to radio coverage. GAMA is funded by the STFC (UK) , the ARC (Australia), the AAO, and the participating institutions. The GAMA website is: http://www.gama-survey.org. \newline
The Herschel-ATLAS is a project with Herschel, which is an ESA space observatory with science instruments provided by European-led Principal Investigator consortia and with important participation from NASA. The H-ATLAS website is http://www.h-atlas.org.\newline
GALEX (Galaxy Evolution Explorer) is a NASA Small Explorer, launched in April 2003. We gratefully acknowledge NASA's support for construction, operation, and science analysis for the GALEX mission, developed in cooperation with the Centre National d'Etudes Spatiales (CNES) of France and the Korean Ministry of Science and Technology.\newline
RJT thanks Wolfgang Kr\"atschmer for discussions on the formation and growth of interstellar grains.
MWG acknowledges the support of the International Max-Planck Research School on Astronomy and Astrophysics Heidelberg (IMPRS-HD).

\appendix

\section{Appendix A: The relation between disk opacity, FIR/submm flux density, and disk scale length in terms of the PT11 model}
\label{AppendixTAUBF2}
In the PT11 model, the opacity of the disk of a spiral galaxy is determined by the mass distribution of the diffuse dust component. This is modeled as the sum of two exponential disks and has been constrained by the reproducible trends found in the radiation transfer analysis of the galaxy sample of \citet{XILOURIS1999}. For such an axisymmetric distribution of diffuse dust the face-on optical depth $\tau_{\nu}(r)$ at a given frequency $\nu$ and a given radial position $r$ is related to the dust surface density $\Sigma(r)$ and the dust spectral emissivity $\kappa_{\nu}$ as $\tau_{\nu}(r) = \Sigma(r)\kappa_{\nu}$. Accordingly, for each disk $i$ in the PT11 model, $\tau_{\nu,i}(r)$ (the face-on optical depth at frequency $\nu$ and radial position $r$ of the disk $i$) can be expressed as:
\begin{equation}
\tau_{\nu,i}(r) = \Sigma_{0,i} \kappa_{\mathrm{ref}} f(\nu)exp\left(\frac{- r}{r_{s,d,\mathrm{ref},i}}\right) = \tau_{0,ref}f(\nu)exp\left(\frac{- r}{r_{s,d,\mathrm{ref},i}}\right)\,,
\end{equation}
where $\Sigma_{0,i}$ is the central dust surface density of the disk $i$, $r_{s,d,\mathrm{ref},i}$ is the scale length of the disk $i$ at a reference wavelength, $\kappa_{\mathrm{ref}}$ is the dust emissivity at a reference frequency, and $f(\nu)$ describes the frequency dependence of the dust emissivity given by the \citet{WEINGARTNER2001} dust model ($f(\nu)$ is not analytically known).
Clearly, in this model geometry, the value of $\Sigma_{0,i}$ is proportional to the mass of dust in the disk $i$ and inversely proportional to the area of the disk, respectively the scale length squared, i.e:
\begin{equation}
\Sigma_{0,i} \propto \frac{M_{\mathrm{dust}}}{r_{s,d,ref,i}^2}\,.
\end{equation}  
Thus, with the opacity of the PT11 model consisting of the sum of two such exponential disks, the optical depth at a given wavelength and position can be fully expressed in terms of the central face-on density of dust, respectively the face-on opacity in a reference band (the B-band at 4430{\AA} for PT11) as: 
\begin{equation}
\tau^f_B = K\frac{M^{\mathrm{diff}}_{\mathrm{dust}}}{r_{s,d,B}^2}\,.
\label{APtaubf1diff}
\end{equation}
where $K = 1.0089\,\mathrm{pc}^2\,\mathrm{kg}^{-1}$ is a constant containing the details of the geometry and the dust model of \citet{WEINGARTNER2001}.\newline
Following the PT11 model, the total mass of dust $M_{\mathrm{dust}}$ in a galaxy is given by
\begin{equation}
M_{\mathrm{dust}} = M^{\mathrm{diff}}_{\mathrm{dust}} + M^{\mathrm{clump}}_{\mathrm{dust}} = (1+\xi) M^{\mathrm{diff}}_{\mathrm{dust}}
\label{APMdiffMF}
\end{equation} 
where $M^{\mathrm{diff}}_{\mathrm{dust}}$ is the mass of diffusely distributed dust and $M^{\mathrm{clump}}_{\mathrm{dust}}$ is the mass of dust in self-shielded clumps, not partaking in the attenuation of optical emission. PT11 find the mass fraction of these clumps to be low ($\sim10-15$\%\footnote{This assumes the emissivity of dust in clumps is the same as that of diffuse dust. As cold self-shielded environments are conducive to the formation of ices with greater emissivity coefficients, this estimate likely represents a upper bound}). Given the uncertainties on the measurement of dust masses in comparison to the likely value of $\xi$, $\xi$ may be neglected to obtain:
\begin{equation}
\tau^f_B = K\frac{M^{\mathrm{diff}}_{\mathrm{dust}}}{r_{s,d,B}^2} \approx K\frac{M_{\mathrm{dust}}}{r_{s,d,B}^2} \,,
\end{equation}
i.e. Eq.~\ref{taubf1} in \S \ref{RTmodel}.\newline

Estimating $\tau^{f}_{B}$ from observable quantities requires several assumptions in order to re-express Eq.~\ref{APtaubf1diff} in terms of observables. 
Under the assumption that the dust emission in the FIR, i.e. at wavelengths longwards of $100\,\mu$m, can be approximated by a modified Planckian with emissivity $\beta$, the total mass of dust in the galaxy can be expressed as:
\begin{equation}
M_{\mathrm{dust}} = \frac{L_{\nu}(\nu_{\mathrm{em}})}{4 \pi \kappa_{\nu_{\mathrm{cal}}}  \left(\frac{\nu_{\mathrm{em}}}{\nu_{\mathrm{cal}}}\right)^{\beta} B(\nu_{\mathrm{em}} ,T_0)}=\frac{S_{\nu}(\nu_{\mathrm{ob}}) D_{\mathrm{L}}^2(z) \nu_{\mathrm{cal}}^{\beta}}{(1+z)^{1+\beta} \kappa_{\nu_{\mathrm{cal}}} \nu_{\mathrm{ob}}^{\beta} B((1+z)\nu_{\mathrm{ob}},T_0)}\,,
\label{APMdust}
\end{equation} 
where $L_{\nu}(\nu_{\mathrm{em}})$ is the luminosity density at the frequency $\nu_{\mathrm{em}}$ related to the observed frequency $\nu_{\mathrm{ob}}$ as $\nu_{\mathrm{ob}} = (1+z)\nu_{\mathrm{em}}$, $S_{\nu}(\nu_{\mathrm{ob}})$ is the observed flux density, $\kappa_{\nu_{\mathrm{cal}}}$ is the emissivity coefficient at the frequency $\nu_{\mathrm{cal}}$, $B(\nu,T)$ is the Planck function evaluated at frequency $\nu$ and temperature $T$, $T_0$ is the restframe temperature of the source, $z$ is the redshift of the source, and $D_{\mathrm{L}}(z)$ is the source's luminosity distance.\newline
Similarly, the physical scale-length of the disk $r_{s,d,B}$ can be expressed as an angular size $\theta_{s,d,B}$ as:
\begin{equation}
r_{s,d,B} = \theta_{s,d,B} D_{\mathrm{A}}(z) = \theta_{s,d,B} \frac{D_{\mathrm{L}}(z)}{(1+z)^2}\,,
\label{APangsize}
\end{equation}
where $D_{\mathrm{A}}(z)$ is the angular diameter distance corresponding to the redshift $z$. Using Eqs.~\ref{APMdiffMF},\ref{APMdust}, and \ref{APangsize} Eq.~\ref{APtaubf1diff} can be expressed as:
 
\begin{eqnarray}
\tau^{f}_{B} &=& \frac{K}{(1+\xi)}\frac{S_{\nu}(\nu_{\mathrm{ob}}) D_{\mathrm{L}}^2(z) \nu_{\mathrm{cal}}^{\beta}}{(1+z)^{1+\beta} \kappa_{\nu_{\mathrm{cal}}} \nu_{\mathrm{ob}}^{\beta} B((1+z)\nu_{\mathrm{ob}},T_0)} \frac{(1+z)^4}{\theta_{s,d,B}^2D_{\mathrm{L}}^2(z)} \nonumber \\
 &=& \frac{K}{(1+\xi)\kappa_{\nu_{\mathrm{cal}}} \nu_{\mathrm{cal}}^{-\beta}\nu_{\mathrm{ob}}^{\beta}\gamma^2} \frac{(1+z)^{3-\beta}}{B((1+z)\nu_{\mathrm{ob}},T_{0})}\frac{S_{\nu}(\nu_{\mathrm{ob}})}{\theta_{s,d,r}^2} \nonumber \\
 &=& A\frac{(1+z)^{3-\beta}}{B((1+z)\nu_{\mathrm{ob}},T_{0})}\frac{S_{\nu}(\nu_{\mathrm{ob}})}{\theta_{s,d,r}^2}\,,
 \label{APstep2}
\end{eqnarray}
corresponding to Eq.~\ref{taubf2} in \S \ref{RTmodel}, with the fixed geometry of the PT11 model being used to re-express $r_{s,d,B}$ as $r_{s,d,B} = \gamma r_{s,d,r}$ (although we set $\xi=0$ in the work presented here for the purpose of determining dust masses, we have chosen to include the factor $(1+\xi)$ in the derivation presented here for purposes of completeness). \newline
Although the approximation of the dust emission from a galaxy by a single temperature modified Planckian is a reasonable assumption at FIR/submmm wavelengths, real galaxies will tend to have a range of components of different temperatures and the temperature derived will correspond to a luminosity weighted average temperature. Furthermore, the emissivity of the dust model of \citet{WEINGARTNER2001} is only approximately a modified Planckian with a fixed emissivity $\beta$, and the actual mass fraction of dust in clumps is not known and difficult to constrain, as the emissivity in these regions may vary with respect to that in the diffuse medium.\newline
Here we have attempted to take these effects into account in first oder by empirically determining the numerical value of $A$ using the radiation transfer solutions to the \citet{XILOURIS1999} galaxy sample, in particular NGC891. For a known source with $\tau^f_B = \tau^f_{B,\mathrm{ref}}$, $\theta_{r,s,r,\mathrm{ref}} = \theta^{\mathrm{ref}}_{r,s,r}$, $S_{\nu}(\nu_{\mathrm{ob}})=S^{\mathrm{ref}}_{\nu}(\nu_{\mathrm{ob}})$, $z=z_{\mathrm{ref}}$, and $T_{0}=T^{\mathrm{ref}}_{0}$ Eq.~\ref{APstep2} can be used to identify $A$ as:
\begin{equation}
A =  \tau^f_{B,\mathrm{ref}} \frac{\theta_{r,s,r,\mathrm{ref}}^2}{S^{\mathrm{ref}}_{\nu}(\nu_{\mathrm{ob}})}\frac{B((1+z_{\mathrm{ref}})\nu_{\mathrm{ob}},T^{\mathrm{ref}}_{0})} {(1+z_{\mathrm{ref}})^{3-\beta}}\,. 
\end{equation}
From the analysis of the \citet{XILOURIS1999} galaxy sample, in particular NGC891 as presented in PT11, we obtain $A = 6.939 \cdot 10^{-13}\,$~$\mathrm{arcsec}^2\,$~$\mathrm{J}\,$~$\mathrm{Jy}^{-1}\,$~$\mathrm{s}^{-1}\,$~$\textrm{Hz}^{-1}\,$~$\textrm{m}^{-2}\,$~$\textrm{ster}^{-1}$ using $\lambda_{\mathrm{ob}} = 250\,\mu$m, $\tau^f_{B,\mathrm{ref}} = 3.5$,  $\theta_{r,s,r,\mathrm{ref}} = 116$", $S^{\mathrm{ref}}_{\nu}(\nu_{250}) = 115\,$Jy, and $T^{\mathrm{ref}}_{0}=20.74\,$K at a distance of 9.5 Mpc.\newline
This empirical calibration implicitly takes the mass fraction of clumps as assumed in the PT11 model into account, hence derived dust masses may be expected to be slightly underestimated ($\lesssim10$\%). \newline

Finally we wish to draw attention to the fact that the grain absorption cross sections in the UV and FIR of the \citet{WEINGARTNER2001} model have been empirically constrained with respect to the hydrogen gas column through measurements of extinction and emission of diffuse dust in the Milky Way. Thus the values of opacity are empirically constrained per unit hydrogen column, i.e $\kappa_{\nu}=\kappa_{\nu,\mathrm{H}}$.  The derivation of dust masses, as e.g. given above, however, requires the absorption cross sections to be expressed per unit grain mass, i.e $\kappa_{\nu}=\kappa_{\nu,\mathrm{m}}$. With the conversion unit hydrogen column to unit grain mass being relatively uncertain, the cross sections in the UV and FIR/submm are much more tightly constrained with respect to each other, than their absolute values. For example, as noted in \citet{DRAINE2007}, the value of $\kappa_{\nu,\mathrm{m}}$ for the model of \citet{WEINGARTNER2001} requires more heavy elements than are expected to be available and may easily overestimate the mass of dust by a factor of $\sim1.4$. In terms of the analysis presented here, such an overestimate will only affect absolute values such as dust masses, while leaving the predicted attenuations unaffected.\newline

\section{Appendix B: The relation between apparent and intrinsic sizes}
\label{AppendixSizes}
\begin{figure}
\plotone{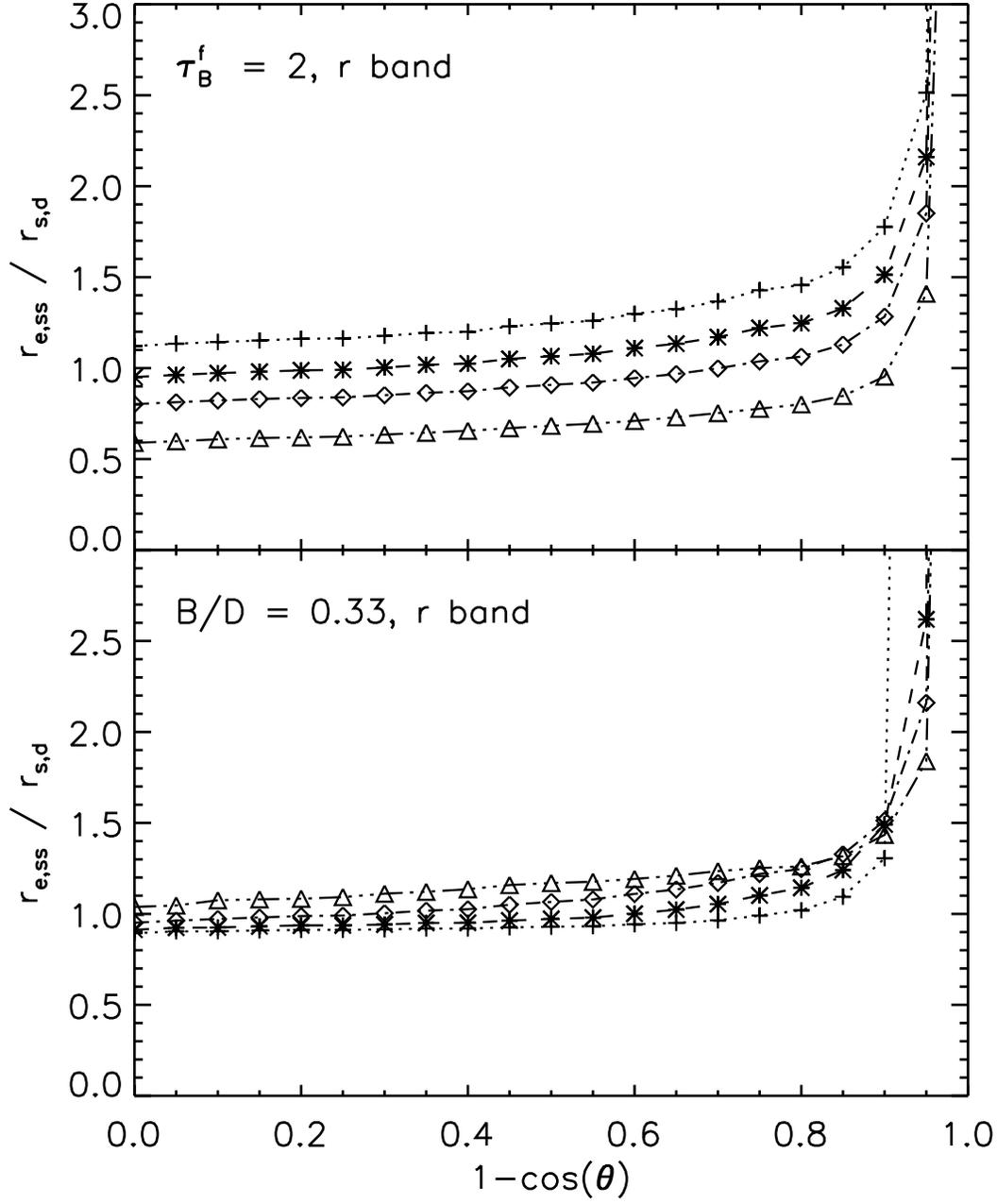}
\caption{Top: Ratio between effective radius observed fitting a single S\'ersic profile ($r_{e,ss}$) and the scale-length of the disk component ($r_{s,d}$) as a function of inclination, for four values of $B/D$; 0.25 (dotted crosses), 0.33 (dashed stars), 0.4 (dash-dotted diamonds), and 0.5 (triple-dash-dotted triangles) observed in the $r$ band.
Bottom: Ratio between effective radius observed fitting a single S\'ersic profile ($r^{ss}_e$) and the scale-length of the disk component ($r^{d}_s$) as a function of inclination, for four values of $\tau^f_B$ at $B/D=0.33$;  0.5 (dotted crosses), 1.0 (dashed stars), 2.0 (dash-dotted diamonds), and 4.0 (triple-dash-dotted triangles) observed in the $r$ band.
Notice the ratio of order unity for values of $B/D$ corresponding to (massive) spiral galaxies, rather than 1.68 as expected for pure disk systems. Data from Pastrav et al. (in prep.)}
\label{Fig_Bogdancorrections}
\end{figure}

Spiral galaxies are fundamentally multi-component systems, consisting, to first order, of a disk and a bulge. Nevertheless, their light profiles are often fitted using single S\'ersic profiles, especially in the case of marginally resolved systems. In order to link the observed sizes, i.e the effective radius, to the intrinsic length scales of the disk and the bulge, multiple factors must be considered.\newline
While the ratio between effective radius and scale-length  for a simple exponential disk is 1.68, the ratio between the effective radius of a single S\'ersic profile fit to a bulge + disk system and the scale-length of the disk component will decrease as the importance of the bulge increases.\newline 
Conversely, the presence of diffuse dust in a late-type galaxy will influence the measured sizes of these objects if the surface density of diffuse dust possesses a radial gradient. Under these circumstances the apparent size measured will tend to be larger than the intrinsic size. The severity of this effect depends on both the value of $\tau^f_B$ and the inclination of the disk $i_d$, and is sensitive to the details of the dust geometry in the galaxy. Furthermore, as the degree of attenuation caused by diffuse dust varies as a function of wavelength, the effect will also be wavelength dependent. This effect has been quantitatively predicted for pure disk systems \citep{MOELLENHOFF2006,PASTRAV2013}, and has been observed in the wavelength dependence of galaxy sizes (e.g., \citealt{KELVIN2011,HAEUSSLER2012}).\newline
A joint consideration of these effects, investigating the combined dependencies of the ratio between observed single S\'ersic effective radius and and the scale-length of the disk component on wavelength, bulge-to-disk ratio $B/D$, inclination $i_d$, and $\tau^f_B$ has been performed by Pastrav et al. (in prep.). Fig.~\ref{Fig_Bogdancorrections} shows the $r$ band ratio as a function of inclination for 4 values of $B/D$ at fixed $\tau^f_B$ (top) and 4 values of $\tau^f_B$ at fixed $B/D$ (bottom). Pastrav et al., have performed this analysis using synthetic images of galaxies created using the same geometry assumed in PT11 and, accordingly, the use of these corrections is entirely consistent with the use of the PT11 radiation transfer model.\newline

\subsection{Determining $r_{d,s}$ and $\tau^f_B$}
Eq.~\ref{taubf2} enables the determination of $\tau^f_B$ based on the observed FIR flux $S_{250}$ and the angular size corresponding to the disk scale-length in the $r$ band $\theta_{s,d,r}$. Taking the corrections into account Eq.~\ref{taubf2} can be expressed as
\begin{equation}
\tau^f_B \propto \frac{S_{250}}{\theta_{e,ss,r}} R(\tau^f_B, i_d, B/D)\;,
\label{eq_ittau}
\end{equation}
 
where $R(\tau^f_B, i_d, B/D)$ is the inverse of the ratio between $r_{e,ss,r}$ (the physical effective radius in the $r$ band obtained from the single S\'ersic fit) and $r_{s,d,r}$ as derived using the ratios of  Pastrav et al.
These ratios are provided in tabulated form, and we have interpolated the ratios in $\lambda$,$i_d$, and $B/D$ and have fit the $\tau^f_B$ dependence using a cubic spline. Using this spline Eq.~\ref{eq_ittau} is solved numerically, obtaining the values of $\tau^f_B$ and $r_{s,d,r}$ for the galaxy.\newline  
In determining the values of $\tau^f_B$ for the \textit{OPTICAL+FIR} sample we have used a value of $B/D = 0.33$, representative of the massive spirals in the sample \citep{GRAHAM2008}. We caution, however, that the value of $B/D$ is a major source of uncertainty, which will be addressed in future work as and when higher resolution imaging, enabling morphological decompositions of the bulge + disk, becomes available.

\end{document}